\documentclass[sn-mathphys-num]{sn-jnl}

\usepackage{amssymb}
\usepackage{amstext}
\usepackage{amsmath}
\usepackage[table]{xcolor}
\usepackage{color}
\usepackage{lscape}
\usepackage{multirow}
\usepackage{arydshln}
\usepackage{url,floatflt}
\usepackage{subfig}
\usepackage{rotating}
\usepackage{amsfonts}
\usepackage{cancel}
\usepackage{amssymb}
\usepackage{stackrel}
\usepackage{tikz}

\usepackage{todonotes}

\definecolor{blue}{RGB}{195,128,31}

\newcommand{\Zset}{\mathbb{Z}}
\newcommand{\Fset}{\mathbb{F}}

\newcommand{\card}[1]{\left|#1\right|}
\renewcommand{\vec}[1]{\text{\mathversion{bold}$#1$}}

\newcommand{\bi}[2]{\binom{#1}{#2}}
\usepackage{arydshln}
\usepackage{bm}
\usepackage{lscape}

\usepackage[amsmath,thmmarks]{ntheorem}
\theoremheaderfont{\bf}
\theorembodyfont{\itshape}
\theoremstyle{plain}
\theoremseparator{:}
\theoremsymbol{}
\theoremstyle{thmstyleone}%
\newtheorem{theorem}{Theorem}
\newtheorem{example}{Example}%

\theoremstyle{thmstylethree}%
\newtheorem{definition}{Definition}%

\newtheorem{lemma}{Lemma}
\newtheorem{corollary}{Corollary}

\theoremheaderfont{\sc}
\theorembodyfont{\upshape}
\theoremstyle{nonumberplain}
\theoremsymbol{\ensuremath{\square}}
\newtheorem{proof}{Proof}

 \usepackage[symbol]{footmisc}

\usepackage{tikz}

\usepackage{multirow}

\newcommand{\supp}[1]{\qopname\relax o{Supp}\left(#1\right)}
\renewcommand{\vec}[1]{\text{\mathversion{bold}$#1$}}

\begin{document}





\title{A New Representation of Binary Sequences by means of Boolean Functions}

\newcommand{\orcidauthorA}{0000-0003-0225-5106} 
\newcommand{\orcidauthorB}{0000-0002-1497-6456} 
\newcommand{\orcidauthorC}{0000-0002-8261-3550}  



\author[1]{ \sur{Miguel Beltr\'{a}}}\email{miguel.beltra@es}

\author[2]{\sur{Sara D. Cardell}}\email{sd.cardell@unesp.br}

\author[3]{ \sur{Amparo F\'uster-Sabater}}\email{amparo.fuster@csic.es}

\author*[1]{ \sur{Ver\'onica Requena}}\email{vrequena@ua.es}

\affil*[1]{\orgdiv{Departamento de Matemáticas}, \orgname{ Universidad de Alicante}, \orgaddress{\city{Alicante},  \country{Spain}}}

\affil[2]{\orgdiv{Departamento de Matemática}, \orgname{Instituto de Geociências e Ciências Exatas, Unesp}, \orgaddress{\street{Campus Rio Claro}, \country{ Brazil}}}

\affil[3]{\orgdiv{Instituto de Tecnolog\'ias F\'isicas y de la Informaci\'on}, \orgname{Consejo Superior de Investigaciones Cient\'ificas}, \orgaddress{\city{Madrid},  \country{Spain}}}


%
%

\abstract{
Boolean functions and binary sequences are main tools used in cryptography. 
In this work, we introduce a new bijection between the set of Boolean functions and the set of binary sequences with period a power of two. We establish a  connection between them which allows us to study some properties of Boolean functions through binary sequences and vice versa. Then, we define a new representation of sequences, based on Boolean functions and derived from the algebraic normal form, named reverse-ANF. Next, we study the relation between such a representation and other representations of Boolean functions as well as between such a representation and the binary sequences. Finally, we analyse the generalized self-shrinking sequences in terms of Boolean functions and some of their properties using the different representations.
}







\maketitle
\section{Introduction}

Boolean functions have applications in propositional logic, electrical engineering, game
theory, reliability, combinatorics, and linear programming \cite{Crama2011}.
They are also very important in complexity theory  \cite{Krause2011,Papadimitriou1994},
coding theory and cryptography \cite{MacWilliams1988bk, Carlet2021}, social sciences  \cite{Ragin1987,Feldman2000},
medicine  and 
biology   \cite{Bonates2006,Shmulevich2002}.
Boolean functions are a useful tool for building models inside a large number of processes in nature, logic, engineering, or science.
In particular, in \textit{cryptography} they are used intensively in the construction and analysis of cryptosystems.
They play an important role in block ciphers, stream ciphers, and hash functions \cite{Braeken2004, Meier1990}.
For example, the implementation of a substitution box (or S-box) needs non-linear Boolean functions to resist cryptanalytic attacks such as differential cryptanalysis \cite{Adams1997,Gupta2005,Nyberg1991b}.
%
In the most common models of stream ciphers, the keystream is produced by means of Boolean functions \cite{AlShehhi2011,Wei2007,Zhang2000}. One
of their basic requirements 
is that they allow to increase the linear complexity \cite{Meier1990,Rueppel1986,Rueppel1987}, provided that such
functions have a high algebraic degree 
\cite{Fuster2014}.
A variety of criteria for the selection of Boolean functions define their abilities to provide security and thus to be used in different applications.

Cryptographic random bit generators are often constructed by
combining several maximum-length LFSRs (Linear Feedback Shift Registers \cite{Golomb1982bk}) in a nonlinear way. 
The
output bits of different LFSRs are combined  using  a nonlinear Boolean function to
get the keystream sequence \cite{Chowdhury2003,Sadkhan2017}. 
In order to eliminate the predictability of
the keystream, the Boolean function should have several properties such as
high algebraic degree, high non-linearity and high correlation immunity \cite{Meier1990,Siegenthaler1984,Wagner2008MT}. 
 
Binary sequences with a period power of two can be represented by the binomial representation introduced in 
\cite{Cardell2019a}. This representation allows to analyse certain cryptographic properties of these sequences such as linear complexity and period \cite{Cardell2020c,Cardell2019a}. 
Some studies relate Boolean functions and binary sequences; see, for example, \cite{Pirsic2012}.
In this work, authors consider the truth table of a Boolean function as a binary sequence, and the polynomial representation used is the ANF (Algebraic Normal Form). They focus their study in the relation of a pseudorandomness measure (correlation measure of order k) for these sequences and two
quality measures (sparsity and combinatorial complexity) for the corresponding Boolean functions. Furthermore, they apply their results to Sidelnikov sequences. In our work, we make use of the binomial representation of a binary sequence and its reveser-ANF to research other cryptographic properties; we do not focus on the pseudorandomness measures of these binary sequences. We examine more general properties such as: algebraic degree, balancedness or linear complexity, among others; moreover, we apply these results to the family of generalized self-shrunken sequences.

In this work, we define a new bijection between the set of Boolean functions (with any number of variables) and the set of binary sequences of period a power of two. For that purpose, we have introduced a new representation, for Boolean functions, based on the Algebraic Normal form, called the reverse-ANF (r-ANF). This new representation facilitates us the link with the binary sequences. There exist different ways to represent a Boolean function (for more details, see \cite{Carlet2021}). However, we have found it necessary to define this new Boolean representation because using other existing representations in the literature as, for instance, the ANF, the bijection defined fails. The binary sequence obtained is not unique, it depends on the number of variables of the ANF. This justification will be explain in more detail in the subsection~\ref{sec:NewRep}.
The aim of this connection is to analyse  properties of Boolean functions through binary sequences and vice versa.
We focus our study on the reverse-ANF of the generalized self-shrunken sequences; since that
they have many interesting cryptographic properties\cite{Cardell2022,Cardell2020c,Hu2004}, and some open problems  could be solved using Boolean functions.
As far as we know, there are no works that analyse these sequences through Boolean functions. 


This paper is organized as follows.
In Section~\ref{sec:Prel}, we present the main concepts, related to Boolean functions and binary sequences, needed to understand the rest of the paper.
Next, in Section~\ref{BR_ANF},  
 we analyse the relation between between the binomial representation 
  and the algebraic normal form of a sequence. We observe that it coincides with the known relation between the truth table and the ANF of a Boolean function. 
After that, in Section~\ref{sec:NewRep} we introduce the new representation of a binary sequence, the called reverse-ANF, and study the existing relation between such a representation and the rest of representations previously introduced. Moreover, we expose some results about the Sierpinski triangle in terms of the r-ANF. In Section~\ref{reverse}, we define the reverse sequence of a given sequence and analyse its connection with shifted versions of the original sequence. The r-ANF of the family of generalized sequences is examined in Section~\ref{sec:gen}. 
Finally, the paper closes in Section~\ref{sec:Concl} with some conclusions and future work. 
\color{black}
 \section{Preliminaries} \label{sec:Prel}
In this section, we introduce the necessary concepts to understand the rest of the paper. In the following subsections, 
we present some important results concerning binary sequences and Boolean functions which help us to define a new connection between them.

\subsection{Sequences}
 
Let $\mathbb{F}_2=\{0,1\}$ be the Galois field of two elements.
Consider $\{u_\tau\}_{\tau\geq 0}=\{u_0, u_1, u_2, \ldots\}$ a binary sequence with $u_\tau\in \mathbb{F}_2$, for $\tau=0, 1, 2,\ldots$
We say that the sequence $\{u_\tau\}_{\tau\geq 0}$, or simply $\{u_\tau\}$, is periodic if  there exists an integer $T$, called period, such that $u_{\tau+T}=u_\tau$, for all $\tau\geq 0$.
In the sequel, all the sequences considered will be binary sequences and the symbol $+$ will denote the Exclusive-OR (XOR) logic operation.

Let $r$ be a positive integer, and let $a_0, a_1, a_2, \ldots, a_{r-1}$ be constant coefficients with $a_j\in\mathbb{F}_2$, for $j=0,1,\ldots,r-1$.
A binary sequence $\{u_\tau\}$ satisfying the relation
\begin{equation*}\label{eq:1}	
u_{\tau+r} =  a_{r-1} u_{\tau + r-1} + a_{r-2} u_{\tau+r-2} + a_{r-3} u_{\tau+r-3} + \cdots + a_{1} u_{\tau+1} + a_{0} u_{\tau}, \quad \tau\geq 0,
\end{equation*}
is called a ($r$-th order) linear recurring sequence  in $\mathbb{F}_2$.
The terms $\{u_0, u_1, \ldots,$ $u_{r-1}\}$ are referred to as the initial terms of the sequence and define uniquely such a sequence.

The monic polynomial
\[
p(x) =  x^{r} + a_{r-1} x^{r-1} + a_{r-2} x^{r-2} + a_{r-3} x^{r-3} + \cdots + a_{1} x + a_{0}
\in \mathbb{F}_2[x]
\]
is called the characteristic polynomial of the linear recurring sequence and $\{u_\tau\}$ is said to be generated by $p(x)$.

We can generate linear recurring sequences using LFSRs~\cite{Golomb1982bk}.
In fact, an LFSR can be defined as an electronic device with $r$ interconnected memory cells (stages) with binary content. At every clock pulse, the binary element of each stage is shifted to the adjacent stage as well as a new element is computed through the linear feedback to fill the empty stage (see Figure~\ref{fig:LFSR}).
We say that the LFSR has maximal-length if its  characteristic polynomial is primitive. Then, its output sequence is called PN-sequence (Pseudo-Noise sequence) and its period is $T = 2^{r}-1$ (see~\cite{Golomb1982bk}).

\begin{figure}
\caption{LFSR of length $r$\label{fig:LFSR}}
\begin{center}
\begin{tikzpicture}[scale=0.7]
\draw (0,-1.5) node [draw,scale=0.8](R1){\phantom{$u_{\tau+r-1}$}};
\draw (0,-1.5) node [scale=0.8]{$u_{\tau+r-1}$};
\draw (2,-1.5) node [draw,scale=0.8](R2){\phantom{$u_{\tau+r-1}$}};
\draw (2,-1.5) node [scale=0.8]{$u_{\tau+r-2}$};
\draw (4,-1.5) node [ draw,scale=0.8](R3){\phantom{$u_{\tau+r-1}$}};
\draw (4,-1.5) node [scale=0.8]{$u_{\tau+r-3}$};
\draw (5.5,-1.5) node [scale=1.1](puntos2){$\cdots$};
\draw (7,-1.5) node [ draw,scale=0.8](R4){$u_{\tau+1}$};
\draw (9,-1.5) node [ draw,scale=0.8](R5){$u_\tau$};

\draw (0,0) node [draw, circle,scale=0.7](C1){$a_1$};
\draw (2,0) node [draw, circle,scale=0.7](C2){$a_2$};
\draw (4,0) node [draw, circle,scale=0.7](C3){$a_3$};
\draw (5.5,0) node [scale=1.1](puntos){$\cdots$};
\draw (7,0) node [draw, circle,scale=0.7](C4){\phantom{$a_2$}};
\draw (7,0) node [scale=0.6]{$a_{r-1}$};
\draw (9,0) node [,draw, circle,scale=0.7](C5){\phantom{$d_2$}};
\draw (9,0) node [scale=0.7]{$a_r$};

\draw (2,1.5) node [draw,circle, scale=0.5](S1){$+$};
\draw (4,1.5) node [draw,circle, scale=0.5](S2){$+$};
\draw (5.5,1.5) node [scale=1.1](puntos1){$\cdots$};
\draw (7,1.5) node [draw,circle, scale=0.5](S3){$+$};
\draw (9,1.5) node [draw,circle, scale=0.5](S4){$+$};

\draw (C1)--(0,1.5);
\draw [-latex] (C2)--(S1) ;
\draw [-latex] (C3)--(S2) ;
\draw [-latex] (C4)--(S3) ;
\draw [-latex] (C5)--(S4) ;

\draw [-latex] (0,1.5) --(S1);
\draw [-latex] (S1)--(S2);
\draw  (S2) --(puntos1);
\draw [-latex] (puntos1) --(S3);
\draw [-latex] (S3)--(S4);
\draw (S4)--(10,1.5);
\draw (10,1.5)--(10,-1.5);
\draw [-latex](10,-1.5)--(R5);

\draw [-latex] (R5) --(R4);
\draw (R4)--(puntos2);
\draw [-latex] (puntos2) --(R3);
\draw [-latex] (R3)--(R2);
\draw [-latex](R2)--(R1);
\draw [-latex](R1)--(-1.5,-1.5);

\draw [-latex] (R1)--(C1);
\draw [-latex] (R2)--(C2);
\draw [-latex] (R3)--(C3);
\draw [-latex] (R4)--(C4);
\draw [-latex] (R5)--(C5);

\draw (-2.1,-1.5)node[scale=0.8,draw,circle]{$u_{\tau+r}$};

\end{tikzpicture}
\end{center}
\end{figure}

The linear complexity of a sequence $\{u_\tau\}$, denoted by $LC$, is defined as the length of the shortest LFSR that generates such a sequence or, equivalently, as the lowest order linear recurrence relationship that generates such a sequence.
In cryptographic applications, the linear complexity must be as large as possible.
The expected value should be at least half the period (see \cite{Rueppel1986}).

\subsection{Boolean functions} \label{sec:ANF}


A \emph{Boolean function} of $n$ variables is a map of the form $f :\Fset_2^n \longrightarrow \Fset_2$. 
We denote by $\mathcal{B}_{n}$ the set  of all Boolean functions of $n$ variables; it is well known that  $\mathcal{B}_{n}$, with the usual addition of functions, is a linear space of dimension $2^{n}$ over $\Fset_2$.

If  we denote by $\vec{i}$ the coefficient vector of the binary expansion of the integer $i$ in $n$ digits, for $i = 0, 1, 2, \ldots, 2^{n}-1$, then we can represent
$\Fset_{2}^{n} = \{ \vec{i} \ | \ i \in \Zset_{2^{n}} \}$.
The \emph{truth table (TT)} of $f$ (see, for example \cite{olejar1998,pasalic1999}) is the binary sequence of length $2^{n}$ given by
\begin{equation} \label{TT}
  \vec{\xi}_{f}
  =
  \left(f(\vec{0}), f(\vec{1}),\ldots, f(\vec{2^{n}-1})\right).
\end{equation}

Next, we show that the truth table of a Boolean function is perfectly determined from its minterms, fact that will be important for our results. 
A \emph{minterm} of $n$ variables, defined by the vector $\vec{u}=(u_{1}, u_{2}, \ldots, u_{n}) \in \mathbb{F}_{2}^{n}$, is the function  $m_{\vec u} \in \mathcal{B}_{n}$ given by
\begin{equation*}
  m_{\vec{u}}(\vec{x})
    = (1 \oplus u_{1} \oplus x_{1})
        (1 \oplus u_{2} \oplus x_{2})
        \cdots
        (1 \oplus u_{n} \oplus x_{n}),
\end{equation*}
for all $\vec{x}=(x_{1}, x_{2}, \ldots, x_{n}) \in \mathbb{F}_{2}^{n}$.
We use  $m_{\vec{u}}(\vec{x})$ or $m_{u}(\vec{x})$, indistinctly because, as it was previously commented, $\vec{u} \in \mathbb{F}_{2}^{n}$ is the binary expansion of $u \in \Zset_{2^{n}}$.
For $i = 0,1,2,\ldots,2^n-1$, it is evident that $m_{i}(\vec{x}) = 1$ if and only if  $\vec{x} = \vec{i}$.
Therefore, the truth table  
\[
  (m_{i}(\vec{0}), \ m_{i}(\vec{1}), \ \ldots, \ m_{i}(\vec{2^{n}-1})) 
\]
of the function $m_{i}(\vec{x})$ has $1$ in the $i$-th position and $0$ in the remaining.
As a consequence, we have that
\begin{equation*} \label{Prel.eq0}
  \bigoplus_{i=0}^{2^{n}-1} m_{i}(\vec{x}) = 1, \quad \text{for all} \quad \vec{x} \in \mathbb{F}_{2}^{n}.
\end{equation*}
Furthermore,  $m_{i}(\vec{x})=m_{j}(\vec{x})$ if and only if $i=j$, so we can identify the minterm  $m_{i}(\vec{x})$ with the integer $i$ (or the vector $\vec{i}$ as required).

Now, for all $f \in \mathcal{B}_{n}$ is easy to check that
\begin{equation} \label{Prel.eq2x}
  f(\vec{x})
  =
  \bigoplus_{i=0}^{2^{n}-1} f(\vec{i}) \ m_{i}(\vec{x})
\end{equation}
and, as
\[
  \bigoplus_{i=0}^{2^{n}-1}a_{i} \ m_{i}(\vec{x})=0 \quad \text{implies that} \quad a_{i}=0 \quad \text{for} \quad i=0,1,2,\ldots,2^{n}-1,
\]
then, we can state that the set $\{m_{0}, m_{1}, \ldots, m_{2^{n}-1}\}$ is a basis of $\mathcal{B}_{n}$.
Consequently, $\dim{\mathcal{B}_{n}}=2^n$ and, therefore, $\card{\mathcal{B}_{n}}=2^{2^{n}}$.

The \emph{support} of $f$, denoted by $\supp{f}$, is the set of vectors of $\Fset_{2}^{n}$ whose image by  $f$ is $1$, that is, 
\begin{equation} \label{def:supp}
  \supp{f}
  =
  \left\{ \vec{a} \in \Fset_{2}^{n} \ | \ f(\vec{a}) = 1 \right\}.
\end{equation}
Therefore, $\supp{f}$ is composed by the vectors of $\Fset_{2}^{n}$ corresponding to the components of $\vec{\xi}_f$ equal to $1$. Due to the identification given between the elements of $\Fset_{2}^{n}$ and $\Zset_{{2}^n}$, through the binary representation, we can rewrite:
$$
  \supp{f}
  =
  \left\{ i \in \Zset_{2^n} \ | \ f(\vec{i}) = 1 \right\}.
$$
Thus, according to expression~(\ref{Prel.eq2x}), we can express $f(\vec{x})$ as follows:
\begin{equation} \label{minterm}
f(\vec{x})=\bigoplus_{i \in \supp{f}}m_{i}(\vec{x})
\end{equation}
which allows us to identify $\supp{f}$ with the set of indices of the minterms of $f(\vec{x})$.

If $f \in \mathcal{B}_{n}$, then we call \emph{weight} of $f$, denoted by $w(f)$, to the number of $1$s in its truth table. Therefore, $w(f) = \card{\supp{f}}$ 
and, clearly
\[
  w(f)
  =
  \sum_{\vec{a} \in \mathbb{Z}_{2}^{n}} f(\vec{a}).
\]
We said that $f$ is \emph{balanced} if $w(f) = 2^{n-1}$.


There exist several ways to represent a Boolean function (see \cite{Carlet2021}). 
For example, we can write $f(\vec{x})$ uniquely
(see \cite{olejar1998,pasalic1999}) as
\begin{equation} \label{eq6}
  f(\vec{x}) = f(x_1,x_2,\ldots,x_n)= \sum_{\vec{u} \in \Fset_{2}^{n}} \mu_{f}(\vec{u}) \vec{x}^{\vec{u}},
\end{equation}
where $\mu_{f}(\vec{u}) \in \Fset_{2}$ is also a Boolean function, 
called the M\"{o}bius transform of $f$ (see, \cite{Rota1964}). If $\vec{u} = (u_{1}, u_{2}, \ldots, u_{n})$, then
\begin{equation*} \label{eq:xANF}
  \vec{x}^{\vec{u}} = x_{1}^{u_{1}} x_{2}^{u_{2}} \cdots x_{n}^{u_{n}}
  \quad \text{with} \quad
  x_{j}^{u_{j}}
  =
  \begin{cases}
    x_{j}, & \text{if $u_{j} = 1$}, \\
    1,     & \text{if $u_{j} = 0$}.
  \end{cases}
\end{equation*}
Observe that in equation~(\ref{eq6})  each   one of the terms $\vec{x}^{\vec{u}}$ is a monomial whose degree is $w(\vec{u})$, corresponding to the number of variables appearing in the product. Note that $w(\vec{u})$ is the number of nonzero components of the vector $\vec{u}$.
This representation of $f$ is known as the \textit{algebraic normal form} (ANF) of $f(\vec{x})$ \cite{Crama2011}.
Denote by $\vec{m}_{\mu_f}$ the vector composed by the coefficients of the ANF of $f(\vec{x})$ given in expression~(\ref{eq6}), that is,
\begin{equation}\label{vec:ANF}
   \vec{m}_{\mu_f} = \left( \mu_{f}(\vec{0}), \mu_{f}(\vec{1}), \ldots, \mu_{f}(\vec{2^n-1})\right).
\end{equation}

\color{black}

We call \emph{degree} of $f(\vec{x})$, denoted by $\deg{f}$, to the maximum of degrees of the monomials of its ANF, that is
\[
  \deg{f}
  =
  \max \{ w(\vec{u}) \ | \  \mu_{f}(\vec{u}) = 1 \}.
\]

\subsection{Relation between the TT and  the ANF} \label{sub:TT_ANF}

In this section, we summarise the well-known relation between the truth table and the ANF of $f$ using  Hadamard matrices (see \cite{Pieprzyk2011}).
This connection will be crucial in order to understand the new Boolean representation  for binary sequences named reverse-ANF and introduced later in Section~\ref{sec:NewRep}. 
\color{black}

The transformation of $f$ into its ANF can be performed by using its corresponding truth table (see \cite{Carlet2021,Pieprzyk2011}).
We call \emph{binomial matrix} to the binary Hadamard matrix of size $2^{t} \times 2^{t}$ defined by
	\[
	H_{t}=
	\left[
	\begin{array}{c c}
		H_{t-1} & H_{t-1} \\
		0_{t-1} & H_{t-1}
	\end{array} 
	\right] \quad \text{and} \quad H_0=\left[1 \right],
	\]
with $0_{t-1}$ the null matrix of size $2^{t-1} \times 2^{t-1}$ and $t>0$ a positive integer.
It is well known that the binomial matrix $H_t$ provides a relation between the ANF of a Boolean function and its truth table through the following equation
\begin{equation} \label{eq:ANF2TT}
\vec{\xi}_{f}= \vec{m}_{\mu_f} \cdot H_{t}\; \text{mod}\; 2,
\end{equation}
where the vector $\vec{\xi}_{f}$ corresponds to the truth table of $f(\vec{x})$, given in equation~\eqref{TT}, and $\vec{m}_{\mu_f}$ is the vector associated with the ANF of $f(\vec{x})$, defined in expression\eqref{vec:ANF}.
As the inverse of $H_t$ is the same matrix ($H_t$ is a idempotent matrix), we can rewrite the equation (\ref{eq:ANF2TT}) as
\begin{equation} \label{eq:TT2ANF}
 \vec{\xi}_{f} \cdot H_{t}\; \text{mod}\; 2 = \vec{m}_{\mu_f}.
\end{equation}
The previous expression allows us to compute the coefficients of the ANF in terms of the binomial matrix $H_{t}$ and of the elements of the truth table of $f(\vec{x})$.

Observe that, from expression~(\ref{Prel.eq2x}), we have that a Boolean function $f$ can be expressed as a linear combination of its minterms, that is, $\xi_f$ can be interpreted as the vector of the coefficients of the representation of $f$ from minterms. Moreover, from expressions~(\ref{def:supp}) and~(\ref{minterm}), the $\supp{f}$ is the set of minterms of $f$. It means that we can obtain the support of a Boolean function (equivalently, the set of its minterms) from its ANF, and vice versa.
 
 The following examples clarify this relation.

\begin{example}
 Consider the truth table of a Boolean function of $4$ variables given by 
 \[
 \vec{\xi}_f=(0,0,0,1,1,0,1,1,1,1,1,1,1,0,0,0) 
 \]
 and the Hadamard matrix
  \begin{eqnarray*}\label{eq:matriz_H4}
		H_{4}
		=
		\left[
		\begin{array}{c c c c c c c c |   c c c c c c c c}
			1  &   1  &   1  &   1  &   1  &   1  &   1  &   1 & 1  &   1  &   1  &   1  &   1  &   1  &   1  &   1 \\
			0  &   1  &   0  &   1  &   0  &   1  &   0  &   1 & 0  &   1  &   0  &   1  &   0  &   1  &   0  &   1 \\
			0  &   0  &   1  &   1  &   0  &   0  &   1  &   1 & 0  &   0  &   1  &   1  &   0  &   0  &   1  &   1 \\
			0  &   0  &   0  &   1  &   0  &   0  &   0  &   1 & 0  &   0  &   0  &   1  &   0  &   0  &   0  &   1 \\
			0  &   0  &   0  &   0  &   1  &   1  &   1  &   1 & 0  &   0  &   0  &   0  &   1  &   1  &   1  &   1 \\
			0  &   0  &   0  &   0  &   0  &   1  &   0  &   1 & 0  &   0  &   0  &   0  &   0  &   1  &   0  &   1 \\
			0  &   0  &   0  &   0  &   0  &   0  &   1  &   1 & 0  &   0  &   0  &   0  &   0  &   0  &   1  &   1 \\
			0  &   0  &   0  &   0  &   0  &   0  &   0  &   1 & 0  &   0  &   0  &   0  &   0  &   0  &   0  &   1 \\
			\hline
			0  &   0  &   0  &   0  &   0  &   0  &   0  &   0 & 1  &   1  &   1  &   1  &   1  &   1  &   1  &   1 \\
			0  &   0  &   0  &   0  &   0  &   0  &   0  &   0 & 0  &   1  &   0  &   1  &   0  &   1  &   0  &   1 \\
			0  &   0  &   0  &   0  &   0  &   0  &   0  &   0 & 0  &   0  &   1  &   1  &   0  &   0  &   1  &   1 \\
			0  &   0  &   0  &   0  &   0  &   0  &   0  &   0 & 0  &   0  &   0  &   1  &   0  &   0  &   0  &   1 \\
			0  &   0  &   0  &   0  &   0  &   0  &   0  &   0 & 0  &   0  &   0  &   0  &   1  &   1  &   1  &   1 \\
			0  &   0  &   0  &   0  &   0  &   0  &   0  &   0 & 0  &   0  &   0  &   0  &   0  &   1  &   0  &   1 \\
			0  &   0  &   0  &   0  &   0  &   0  &   0  &   0 & 0  &   0  &   0  &   0  &   0  &   0  &   1  &   1 \\
			0  &   0  &   0  &   0  &   0  &   0  &   0  &   0 & 0  &   0  &   0  &   0  &   0  &   0  &   0  &   1
			
		\end{array}
		\right] .
		\end{eqnarray*}
   We can obtain the ANF from the truth table $\vec{\xi}_f$ by means of equation (\ref{eq:TT2ANF})  as follows
\[
\vec{m}_{\mu_f}=
\vec{\xi}_f \cdot H_{t}\; \text{mod}\; 2 = (0,0,0,1,1,1,0,0,1,0,0,1,1,0,1,1).
\]
In order to obtain the ANF, we need the positions of the $1$s in the vector $\vec{m}_{\mu_f}$ (we assume that this vector starts in position $0$ from left to right). Therefore, we have $1$s in positions $\{3,4,5,8,11,12,14,15\}$, which means that the ANF is given by 
 \begin{align*}
f(\vec{x})& =x^{\vec{3}} + x^{\vec{4}} +x^{\vec{5}}+x^{\vec{8}}+x^{\vec{11}}+x^{\vec{12}}+x^{\vec{14}}+x^{\vec{15}}\\
& =x_1^0x_2^0x_3^1x_4^1+x_1^0x_2^1x_3^0x_4^0+x_1^0x_2^1x_3^0x_4^1+x_1^1x_2^0x_3^0x_4^0+x_1^1x_2^0x_3^1x_4^1+x_1^1x_2^1x_3^0x_4^0+x_1^1x_2^1x_3^1x_4^0+x_1^1x_2^1x_3^1x_4^1\\
&=x_3x_4+x_2+x_2x_4+x_1+x_1x_3x_4+x_1x_2+x_1x_2x_3+x_1x_2x_3x_4
 \end{align*}
 \end{example}
 Next example proceeds in a reverse way.
\begin{example}
 Consider the Boolean function of $4$ variables given by $$f(\vec{x})=f(x_1,x_2,x_3,x_4)=1+x_4+x_2x_3x_4+x_1x_2.$$
 We have that its ANF is $f(\vec{x})=x^{\vec{0}}+x^{\vec{1}}+x^{\vec{7}}+x^{\vec{12}}$ and, therefore, $\vec{m}_{\mu_f}=(1,1,0,0,0,0,0,1,0,0,0,0,1,0,0,0)$.
 Applying equation~(\ref{eq:ANF2TT}), we have that the truth table of $f(\vec{x})$ is
 \[
 \vec{\xi}_{f}= \vec{m}_{\mu_f} \cdot H_{4}\; \text{mod}\; 2 = 
(1,0,1,0,1,0,1,1,1,0,1,0,0,1,0,0),  
\]
and, therefore, we can represent $f(\vec{x})$ through its minterms as follows
\begin{align*}
f(\vec{x})= & m_{0}(\vec{x})+m_{2}(\vec{x})+m_{4}(\vec{x})+m_{6}(\vec{x})+m_{7}(\vec{x}) + m_{8}(\vec{x})+m_{10}(\vec{x})+m_{13}(\vec{x}).
\end{align*}
\end{example}

\subsection{$B$-representation} \label{sec:Brepr}

From now on, we always consider sequences $\{s_\tau\}$ of period a power of $2$.

The binomial representation ($B$-representation) of binary sequences was first introduced in \cite{Cardell2020c,Cardell2019a}. 
In fact, every binary sequence $\{s_\tau\}$ with period   $2^t$  can be expressed as a linear combination of binomial sequences as
\begin{equation}\label{eq:binomial_sumatorio}
b(\{s_\tau\})
=
\sum_{i=0}^{2^t - 1}c_i\left\{\binom{n}{i}\right\},
\end{equation} 
where $c_i\in \Fset_2$ for $i=0,1, \ldots, 2^t - 1$ and $\left\{\binom{n}{i}\right\}$ is the $i$-th binomial sequence \cite{Cardell2019a}. 
This sequence   can be computed as 
\begin{equation*}\label{eq:bin}
\left\{\binom{n}{i}\right\} = \left\{\binom{0}{i}, \binom{1}{i}, \binom{2}{i}, \ldots \right\},
\end{equation*}
where the coefficients $\binom{j}{i}$, for $j=0,1,\ldots$, are reduced modulo $2$ and $\binom{j}{i}=0$ if $i>j$.
If there is no ambiguity, we can   make an abuse of notation and  simply denote the binomial sequence by $\binom{n}{i}$.
In Table~\ref{tab:1}, one can find the first $16$ binomial sequences. 

The  expression~(\ref{eq:binomial_sumatorio}) is called the \emph{B-representation} of a sequence $\{s_\tau\}$, denoted by $b(\{s_\tau\})$.
Sometimes, for simplicity, we will denote the B-representation by 
$
b(\{s_\tau\})
=\left\{
\sum_{i=0}^{2^t - 1}c_i\binom{n}{i}
\right\}$
  or simply 
  $
b(\{s_\tau\})
=
\sum_{i=0}^{2^t - 1}c_i\binom{n}{i}.
 $
 
Consider  $\vec{c}_{\mu_f}$
the vector of length $2^t$ composed by the coefficients of the B-representation of $\{s_\tau\}$, that is, we have that
\begin{equation}\label{vec:cf}
c_i=
\begin{cases}
 1     & \text{if }   \binom{n}{i} \in  b(\{s_\tau\}), \\
0     & \text{if }    \binom{n}{i}  \notin b(\{s_\tau\}),
  \end{cases}
\end{equation} 
 for $i=0,1,\ldots, 2^t-1$.
For convenience, we can also denote the B-representation by the vector $(i_1,i_2, \ldots, i_k)$, where 
$c_{i_j}\not = 0$ in (\ref{vec:cf}), with $(i_1 < i_2< \cdots< i_k)$ for $j=1, \ldots, k$.

Next, we present an example to clarify these two representations.
\begin{example}
Consider the  sequence of period $T=8$, 
$
\{s_\tau\}
=
(1,1,1,0,0,0,0,1)
$
whose  B-representation is given by
$b(\{s_\tau\})=\binom{n}{0}+\binom{n}{3}+\binom{n}{4} $.
We have that $\vec{c}_{\mu_f}=(1,0,0,1,1,0,0,0)$.
Moreover, this sequence can also be represented by the vector $b(\{s_\tau\})=(0,3,4)$.
\end{example}



\begin{table}[h]
\caption{The first 16 binomial coefficients, their binomial sequences $\binom{n}{i}$, periods and complexities} \label{tab:1}
\begin{tabular}{|c|l|c|c|}\hline
\text{Binomial sequence\;\;}&\text{Binary representation} & \text{Period} & LC\\ \hline
$\binom{n}{0}$ & 1 1 1 1 1 1 1 1 1 1 1 1 1 1 1 1\ldots & 1 & 1\\
$\binom{n}{1}$ & 0 1 0 1 0 1 0 1 0 1 0 1 0 1 0 1 \ldots & 2 & 2\\
$\binom{n}{2} $& 0 0 1 1 0 0 1 1 0 0 1 1 0 0 1 1 \ldots & 4 & 3\\
$\binom{n}{3} $& 0 0 0 1 0 0 0 1 0 0 0 1 0 0 0 1 \ldots & 4 & 4\\
$\binom{n}{4} $& 0 0 0 0 1 1 1 1 0 0 0 0 1 1 1 1\ldots & 8 & 5\\
$\binom{n}{5} $& 0 0 0 0 0 1 0 1 0 0 0 0 0 1 0 1 \ldots & 8 & 6\\
$\binom{n}{6} $& 0 0 0 0 0 0 1 1 0 0 0 0 0 0 1 1\ldots & 8 & 7\\
$\binom{n}{7} $& 0 0 0 0 0 0 0 1 0 0 0 0 0 0 0 1\ldots & 8 & 8\\
$\binom{n}{8} $& 0 0 0 0 0 0 0 0 1 1 1 1 1 1 1 1 \ldots & 16 & 9\\
$\binom{n}{9} $& 0 0 0 0 0 0 0 0 0 1 0 1 0 1 0 1 \ldots & 16 & 10\\
$\binom{n}{10}$ & 0 0 0 0 0 0 0 0 0 0 1 1 0 0 1 1 \ldots & 16 & 11\\
$\binom{n}{11}$ & 0 0 0 0 0 0 0 0 0 0 0 1 0 0 0 1 \ldots & 16 & 12\\
$\binom{n}{12}$ & 0 0 0 0 0 0 0 0 0 0 0 0 1 1 1 1 \ldots & 16 & 13\\
$\binom{n}{13}$ & 0 0 0 0 0 0 0 0 0 0 0 0 0 1 0 1 \ldots & 16 & 14\\
$\binom{n}{14}$ & 0 0 0 0 0 0 0 0 0 0 0 0 0 0 1 1 \ldots & 16 & 15\\
$\binom{n}{15}$ & 0 0 0 0 0 0 0 0 0 0 0 0 0 0 0 1 \ldots & 16 & 16\\
\hline
\end{tabular}
\end{table}

The coefficient $c_{i_k}$ and the B-representation  provide us some information about two fundamental parameters of the sequence: period and linear complexity  (see \cite{Cardell2019a}  for more details).
Indeed, note that:
\begin{itemize}
\item The period   of the sequence $\{s_\tau\}$ is the same as that of the binomial sequence $ \binom{n}{i_k} $. 

\item The linear complexity of the sequence $\{s_\tau\}$ is the same as that of the binomial sequence $\binom{n}{i_k}$, that is $
LC = i_k + 1.$
\end{itemize}
Fixed a linear complexity $LC$,  the period  of the sequence is uniquely determined. 
Nevertheless, fixed a period  there exist distinct sequences with such a period but different linear complexities (see for example Table~\ref{tab:1}).


In \cite{Fuster-Sabater2022a}, authors prove that there exists a relation between a binary sequence of period a power of $2$ and its B-representation. This connection is similar to the one between ANF of a Boolean function and its truth table given in expression~(\ref{eq:ANF2TT}). 
 Due to the particular structure of the binomial sequences, the B-representation of a sequence $\{s_\tau\}$ 
can be turned into a matrix equation, as it is showed in the following theorem.
\begin{theorem}[Theorem~2 of \cite{Fuster-Sabater2022a}]\label{th:matriz-Hadamard}
Consider the B-representation of a sequence $\{s_\tau\}$ of period $T=2^t$, with $t$ a non-negative integer, and let $H_{t}$ be the binomial matrix of size $2^{t} \times 2^{t}$. Then,
\begin{equation} \label{eq:matricial}
 (s_0,s_1,\ldots,s_{2^t-1})  = (c_0,c_1,\ldots,c_{2^t-1})\cdot H_{t}\; \text{mod}\; 2,
\end{equation}
  where the vector  $(s_0,s_1,\ldots,s_{2^t-1})$ corresponds to the first $2^t$ terms of the sequence $\{s_\tau\}$, and $(c_0,c_1,\ldots,c_{2^t-1})$ is the vector $\vec{c}_{\mu_f}$ given in expression~(\ref{vec:cf}).
\end{theorem}


As 
the matrix 
$H_t$ is an idempotent matrix, we can write  equation (\ref{eq:matricial}) as
\begin{equation} \label{eq:binomial_matricial}
(s_0,s_1,\ldots,s_{2^t-1}) \cdot H_{t}\; \text{mod}\; 2 = (c_0,c_1,\ldots,c_{2^t-1}).
\end{equation}
Therefore,  the B-representation of a sequence $\{s_\tau\}$ can be computed in terms of the binomial matrix $H_{t}$ and of the elements of the sequence $\{s_\tau\}$, and vice versa.

The following example clarifies this construction.


\begin{example} \label{ex:seqBin}
 Consider the sequence $\{s_\tau\}$ with period $T=16$ given by the vector
 \[
(s_0,s_1,\ldots,s_{15})=(0, 0, 0, 1, 1, 1, 0, 0, 1, 0, 0, 1, 1, 0, 1, 1). 
 \]
 From equation (\ref{eq:binomial_matricial}), we have that 
\[
(c_0,c_1,\ldots,c_{15})=
(s_0,s_1,\ldots,s_{15}) \cdot H_{4}\; \text{mod}\; 2 =
(0,0,0,1,1,0,1,1,1,1,1,1,1,0,0,0),
\]
 and, therefore, the B-representation of $\{s_\tau\}$ is 
 \[
 b(\{s_\tau\})=
\binom{n}{3} +\binom{n}{4}+ \binom{n}{6}+\binom{n}{7}+\binom{n}{8}+\binom{n}{9}+\binom{n}{10}+\binom{n}{11}+\binom{n}{12}.
 \]
 \end{example}

\section{Relation between B-representation and ANF}\label{BR_ANF}

Due to the close relation between expressions~(\ref{eq:TT2ANF}) and (\ref{eq:matricial}), we will provide a connection between binary sequences $\{s_\tau\}$ and Boolean functions. This link will allows us to analyse some parameters of the Boolean functions from the study of the binary sequences, and vice versa.
In particular, in this section, we analyse the relation between the B-representation of a sequence $\{s_\tau\}$ and the ANF of a Boolean function.

An important fact which helps us to understand this relation is that  
the rows of the binomial matrix $H_t$ are the first $2^t$ binomial sequences 
(see, \cite{Fuster-Sabater2022a,Fuster-Sabater2022}), that is, we can express
\[
	H_{t}=
	\left[
	\begin{array}{ccccc}
	\left\{\binom{n}{0}\right\}  & \left\{\binom{n}{1}\right\}  & \cdots   & \left\{\binom{n}{2^{t}-2}\right\}  & \left\{\binom{n}{2^{t}-1}\right\}
	\end{array} 
	\right]^T.
	\]
Therefore, from expression~(\ref{eq:binomial_sumatorio}), we have that every binary sequence $\{s_\tau\}$ with period   $2^t$  can be expressed as a linear combination of the rows of $H_t$.

On the one hand, we have that equation~(\ref{eq:ANF2TT}),  $\xi_f=\vec{m}_{\mu_f} \cdot H_{t}$, 
means that the vector $\xi_f$ is obtained as the linear combination of the rows of $H_t$ which is determined by the elements of $\vec{m}_{\mu_f}$. Therefore, the vector $\xi_f$ is obtained summing up binomial sequences. 
Observe that we can define a relation between minterms and binomial sequences. Moreover, as the support of a Boolean function is the set of its minterms, and the set of the binomial sequences is the B-representation, we can relate both representations. Later, we will go into detail about this idea. 

On the other hand, we have that $\vec{c}_{\mu_f}$, the vector of coefficients of the B-representation of $\{s_\tau\}$, is determined by the coefficients of the linear combination of the binomial sequences 
of $b(\{s_\tau\})$. 
That is, from equation~(\ref{eq:binomial_matricial}) we have that, $\vec{c}_{\mu_f}=\{s_\tau\}\cdot H_t$, and 
we can deduce that the vector  $\vec{c}_{\mu_f}$ is obtained as the linear combination  of the rows of $H_t$ which is determined by the elements of $\{s_\tau\}$.
Moreover, the vector $b(\{s_\tau\})=(i_1,i_2, \ldots, i_k)$ indicates the positions of the $1$s in $\vec{c}_{\mu_f}$, i.e., the binomial sequences which form the B-representation.  Note that this vector is closely related with the concept of the support of a Boolean function.




Therefore, we can deduce that there exists a connection between the vectors $\vec{\xi}_f$ and $\vec{c}_{\mu_f}$, since both provide the B-representation associated to certain sequence. It implies that the ANF of a function $f$, with vector $\vec{m}_{\mu_f}$, and the sequence $\{s_\tau\}$ are related too.
In Figure~\ref{fig:relation}, we show the different connections.

\begin{figure}
\begin{center}
    \begin{tikzpicture}
        \draw (0,0) node {Truth Table = ANF $\cdot$ Hadamard matrix};
                \draw (0,-1) node {B-representation = Sequence $\cdot$ Hadamard matrix};
                \draw [latex-latex](-2.15,-0.25)--(-2.15,-0.8); 
          \draw [latex-latex](-0.25,-0.25)--(-0.25,-0.8); 
                    \draw (-4,-1.30)rectangle (4,0.35);
                    \draw [latex-latex, color=red, thick](-1.75,-0.8)--(-0.75,-0.25); 
    \end{tikzpicture}

\vspace{2em}
       \begin{tikzpicture}
        \draw (0,0) node {$\xi_f = m_{\mu_f} \cdot H_t$};
         \draw (0,-1) node {$c_{\mu_f} =  \{s_\tau\}  \cdot H_t$};
         \draw [latex-latex](-0.85,-0.25)--(-0.85,-0.8); 
          \draw [latex-latex](0.15,-0.25)--(0.15,-0.8); 
          \draw (-1.25,-1.30)rectangle (1.25,0.35);
    \end{tikzpicture}
\end{center}
    \caption{Relation between B-representation and ANF}
    \label{fig:relation}
\end{figure}


From now on, we work indistinctly with the sequence $\{s_\tau\}$ of length $2^t$ (which can be expressed by a vector $\vec{s}$) and $f$ the Boolean function of $t$ variables whose ANF is determined by the vector $\vec{m}_{\mu_f}$, which coincides with the sequence $\vec{s}$. In this case, we say that the sequence $\{s_\tau\}$ is associated with $f$.


 Through these relations, we can state that the set of minterms of a Boolean function $f$ coincides with the vector $b(\{s_\tau\})$ of the B-representation of the sequence $\{s_\tau\}$ associated with $f$.

 The following example shows this relation.
 \begin{example} \label{ex:BR_ANF}
Consider again Example~\ref{ex:seqBin}. We have that $b(\{s_\tau\})=(3,4,6,7,8,9,10,11,12)$.
We can identify this set of indices with the set of minterms of a Boolean function $f(\vec{x})$ of $4$ variables, that is, we have that
\[
\vec{\xi}_f=
(0,0,0,1,1,0,1,1,1,1,1,1,1,0,0,0).
\]
On the one hand, from equation~(\ref{eq:matricial}), we obtain the sequence 
 \[
 (s_0,s_1,\ldots,s_{15})=(0, 0, 0, 1, 1, 1, 0, 0, 1, 0, 0, 1, 1, 0, 1, 1). 
 \]
On the other hand, from equation~(\ref{eq:TT2ANF}), we have that $$\vec{m}_{\mu_f}=(0, 0, 0, 1, 1, 1, 0, 0, 1, 0, 0, 1, 1, 0, 1, 1)=\{s_\tau\}.$$ 
Therefore, the ANF of $f$ is given by $f(\vec{x})=x^{\pmb{3}}+x^{\pmb{4}}+x^{\pmb{5}}+x^{\pmb{8}}+x^{\pmb{11}}+x^{\pmb{12}}+x^{\pmb{14}}+x^{\pmb{15}}$.
We have related a sequence with a Boolean function.
 \end{example}
\section{New representation of sequences}\label{sec:NewRep}

In this section, we introduce a new representation of sequences, named  reverse-ANF, based on the ANF of a Boolean function and the B-representation of such sequences.
We study this representation and analyse its relation with the rest of representations studied in this paper.
\subsection{Reverse-ANF of a sequence}\label{sec:boolean}

In order to introduce this new representation of a sequence, we need to consider $\vec{x}^{\vec{u}}$ in a reverse way (see, for example Table~\ref{tab:pot}), that is, we denote by
\begin{equation}\label{eq:monomio}
  \overline{\vec{x}^{\vec{u} }}= x_{n}^{u_{1}} x_{n-1}^{u_{2}} \cdots x_{1}^{u_{n}}
  \quad \text{with} \quad
  x_{n-j+1}^{u_{j}}
  =
  \begin{cases}
    x_{n-j+1}, & \text{if $u_{j} = 1$}, \\
    1,     & \text{if $u_{j} = 0$}.
  \end{cases}
\end{equation}

\begin{table}[h]
\caption{First 8 powers of $\overline{\pmb{x}^{\vec{u}}}$} \label{tab:pot}
\begin{tabular}{|c|c|c|}\hline
  $\overline{\pmb{x^u}} $& $\pmb{u}$&  $ x_{3}^{u_{1}} x_{2}^{u_{2}} x_{1}^{u_{3}}$ \\\hline\hline
 $\overline{\pmb{x}^\vec{0}}$ & 000& 1  \\\hline
 $\overline{\pmb{x}^\vec{1}} $& 001 & $x_1$ \\\hline
 $ \overline{\pmb{x}^\vec{2}} $& 010& $x_2$  \\\hline
 $  \overline{\pmb{x}^\vec{3}} $& 011& $x_2x_1$  \\\hline
  $   \overline{\pmb{x}^\vec{4}}$ & 100&$x_3$  \\\hline
   $    \overline{\pmb{x}^\vec{5}}$ &101&$ x_3x_1$  \\\hline
    $     \overline{\pmb{x}^\vec{6}}$ &110&$ x_3x_2$  \\\hline
     $     \overline{ \pmb{x}^\vec{7}}$ &111&$ x_3x_2x_1 $ \\\hline
\end{tabular}

\end{table}

Next theorem is the most important result of our paper since that it establishes a bijection between the B-representation of a sequence and a Boolean function.

\begin{theorem} \label{th:bijection}
Let  $\mathcal{S}$ be the set of all binary sequences of period power of two and $\mathcal{B}$ the set of all Boolean functions.
Consider the map given by
\begin{equation}\label{eq:map}
\begin{array}{cccc}
\phi: & \mathcal{S}&\longrightarrow &\mathcal{B}\\
&\sum_{i=0}^{2^t-1}c_i   \binom{n}{i} &\rightarrow &c_0\overline{\pmb{x}^{\pmb{0}}}+c_1\overline{\pmb{x}^{\pmb{1}}}+\cdots+c_{2^t-1}\overline{\pmb{x}^{\pmb{2^t-1}}}\\
\end{array}
\end{equation}
where $\overline{\pmb{x}^{\pmb{i}}}$ is as in equation~(\ref{eq:monomio}), $\pmb{i}$ is the shortest binary representation of $i$ and $t$ represents  any positive integer. Then, $\phi$ is a bijection.
\end{theorem}

\begin{proof}
 

On the one hand, two different sequences in $\mathcal{S}$ have different B-representations, therefore, the vector $(c_0,c_1,\ldots, c_{2^t-1})$ is unique for each sequence. 
Thus, two different sequences in $\mathcal{S}$ correspond to two different Boolean functions in $\mathcal{B}$.

On the other hand, consider a Boolean function $f\in \mathcal{B}$ of $t$ variables, whose 
Boolean representation is 
$$
  f(\vec{x}) = \sum_{i=0}^{2^t-1}c_i \overline{\vec{x}^{\vec{i}}}.
  $$
    This function represents the sequence 
  $$
  \sum_{i=0}^{2^t-1}c_i \left\{\binom{n}{i}\right\}.
  $$
  As a consequence, the map $\phi$ is bijective.
 \end{proof}

\color{black}
Note that, if we had considered the terms $\pmb{x}^{\pmb{j}}$, instead of $\overline{\pmb{x}^{\pmb{j}}}$ in the bijection given in expression~(\ref{eq:map}), then the Boolean representation of a binary sequence would not be unique; since that this would depend on the number of digits considered in the binary representation of~$j$, and, therefore, of the number of variables.
To illustrate this idea, consider the map:
\begin{equation}\label{eq:map2}
\begin{array}{cccc}
\varphi: & \mathcal{S}&\longrightarrow &\mathcal{B}\\
&\sum_{i=0}^{2^t-1}c_i   \binom{n}{i} &\rightarrow &c_0\pmb{x}^{\pmb{0}}+c_1\pmb{x}^{\pmb{1}}+\cdots+c_{2^t-1}\pmb{x}^{\pmb{2^t-1}}\\
\end{array}
\end{equation}
In (\ref{eq:map})
we have that the r-ANF of   $\{s_\tau\}=\left\{\binom{n}{1}\right\}$ is    $\mathcal{B}(\{s_\tau\})=\phi(\{s_\tau\})=\overline{\pmb{x}^{1}}$.
However, the form of the representation in (\ref{eq:map2}) 
 depends on the number of variables. 
If  we consider Boolean functions of 4 variables, the representation in (\ref{eq:map2})   is
$$\varphi(\{s_\tau\})=\pmb{x}^{1}=x_1^0x_2^0x_3^0x_4^1=x_4.$$
Now, if we consider  $5$ variables, then  we have that 
$$\varphi(\{s_\tau\})=\pmb{x}^{1}=x_1^0x_2^0x_3^0x_4^0x_5^1=x_5.$$
Therefore, considering the map given in (\ref{eq:map2}), the same sequence has two different representations, which is not practical for our purposes. 
 For this reason, we consider the bijection $\phi$ given in expression~(\ref{eq:map}) instead of $\varphi$ in expression~(\ref{eq:map2}).
 On that account, we found necessary the introduction of the r-ANF of a sequence ${s_\tau}$ in order to obtain a unique Boolean function representing  the sequence.

 \begin{table}[h]
\caption{Boolean representation of the first $16$ binomial sequences}\label{tab:bin}
\begin{tabular}{|c|c|}\hline
\text{Binomial sequence} & \text{Boolean representation}\\\hline
  $\binom{n}{0}$   & $1$ \\\hline
  $\binom{n}{1} $  &$ x_1$ \\\hline
   $ \binom{n}{2}$   & $ x_2$ \\\hline
  $\binom{n}{3}   $&  $x_2x_1$\\\hline
  $\binom{n}{4}   $& $ x_3$\\\hline
  $\binom{n}{5}   $&$ x_3x_1$ \\\hline
  $\binom{n}{6}   $& $ x_3x_2$\\\hline
  $\binom{n}{7}   $&$ x_3x_2x_1 $\\\hline
  $\binom{n}{8}   $& $x_4$ \\\hline
  $\binom{n}{9}   $&$ x_4x_1$ \\\hline
  $\binom{n}{10}   $&$ x_4x_2$ \\\hline
  $\binom{n}{11}   $&$ x_4x_2x_1$ \\\hline
  $\binom{n}{12}   $&$ x_4x_3$ \\\hline
  $\binom{n}{13}   $&$ x_4x_3x_1$ \\\hline
  $\binom{n}{14}   $&$ x_4x_3x_2 $\\\hline
  $\binom{n}{15}   $&$ x_4x_3x_2x_1$ \\\hline
\end{tabular}
\end{table}

 According to Theorem~\ref{th:bijection}, we can introduce a new representation, in terms of Boolean functions, for binary sequences with period a power of two. 

 \begin{definition} \label{Def:Brep}
 Given a binary sequence $\{s_\tau\}$ with period $2^t$,
 and its B-representation $ b(\{s_\tau\})=\sum_{i=0}^{2^t-1}c_i \binom{n}{i}$,
 the Boolean representation of this sequence, called reverse-ANF of $\{s_\tau\}$, and denoted by $\mathcal{B}(\{s_\tau\})$, is given by $ \mathcal{B}(\{s_\tau\})=\phi(\{s_\tau\})$, where $\phi$ is as in expression~(\ref{eq:map}); that is,
 \begin{equation} \label{rep:Boolean}
 \mathcal{B}(\{s_\tau\})=\sum_{i=0}^{2^t-1}c_i \overline{\vec{x}^{\vec{i}}}.
 \end{equation}

 \end{definition}
 Observe that the 
 vector of the coefficients of the reverse-ANF of $\{s_\tau\}$, given in expression~(\ref{rep:Boolean}), coincides with $\vec{c}_{\mu_f}$, the vector of the coefficients of the B-representation of $\{s_\tau\}$. Therefore, we use the vector $\vec{c}_{\mu_f}$ indistinctly for the B-representation and the r-ANF of a sequence $\{s_\tau\}$.


In Table~\ref{tab:bin} we can find the r-ANF of the first $16$ binomial sequences. 
Note that these r-ANF can be considered as Boolean functions of $4$ variables. Moreover, if we compute their truth tables, we obtain   the binomial sequences of length $16$ given in Table~\ref{tab:1} (evaluating the elements $\vec{x} \in \mathbb{F}_2^{n}$ in reverse order. See Table~\ref{tab:BR} for the case $\binom{n}{1}$).
Next example illustrates this idea. 

\begin{example} \label{ex:rANF}
 Consider the r-ANF  of a sequence $\{s_\tau\}$ of length 
 $2^4$ given by
 \[
\mathcal{B}(\{s_\tau\})=\overline{\pmb{x}^{\pmb{1}}}=x_1.
 \]
The truth table of this Boolean function, obtained in Table~\ref{tab:BR}, is
\[
(0,1,0,1,0,1,0,1,0,1,0,1,0,1,0,1),
\] 
which coincides with the binomial sequence $\binom{n}{1}$ given in Table~\ref{tab:1}. 
\begin{table}[h] 
\caption{Truth table of $\mathcal{B}(\{s_\tau\})=\overline{\pmb{x}^{\pmb{1}}}$}  \label{tab:BR}
\begin{tabular}{|c|c|}\hline
\text{$(x_4,x_3,x_2,x_1)$} & \text{$x_1$}\\\hline
  (0,0,0,0)   & 0\\\hline
  (0,0,0,1)   & 1 \\\hline
  (0,0,1,0)  &  0\\\hline
  (0,0,1,1)   &  1\\\hline
  (0,1,0,0)   &  0\\\hline
  (0,1,0,1)   & 1 \\\hline
  (0,1,1,0)  &  0\\\hline
  (0,1,1,1)  & 1 \\\hline
  (1,0,0,0)   & 0 \\\hline
  (1,0,0,1)   & 1\\\hline
  (1,0,1,0)   & 0 \\\hline
  (1,0,1,1)   & 1\\\hline
  (1,1,0,0)   & 0 \\\hline
  (1,1,0,1)   & 1\\\hline
  (1,1,1,0)   & 0 \\\hline
  (1,1,1,1)   & 1 \\\hline
\end{tabular}
\end{table}
If we want to know the sequence $\{s_{\tau}\}$ associated to this r-ANF, we have to apply the equation~\eqref{eq:matricial}, considering 
$$\vec{c}_{\mu_f}=(0,1,0,0,0,0,0,0,0,0,0,0,0,0,0,0),$$ 
with $\mathcal{B}(\{s_\tau\})=(1)$.
Therefore, we have that
\[
\{s_{\tau}\}=(0,1,0,1,0,1,0,1,0,1,0,1,0,1,0,1),
\]
that is, the truth table of the r-ANF. Moreover, the ANF associated to the sequence $\{s_{\tau}\}$ is
\begin{align*}
f(x_1,x_2,x_3,x_4) & =x^{\vec{1}}+x^{\vec{3}}+x^{\vec{5}}+x^{\vec{7}}+x^{\vec{9}}+x^{\vec{11}}+x^{\vec{13}}+x^{\vec{15}}\\
& = x_4+x_3x_4+x_2x_4+x_2x_3x_4+x_1x_4+x_1x_3x_4+x_1x_2x_4+x_1x_2x_3x_4.
\end{align*}
\end{example}

Observe that in the previous example the ANF associated to the sequence $\{s_{\tau}\}$ and the reverse-ANF are completely different and perform distinct functions. Through the ANF, we obtain the sequence $\{s_{\tau}\}$ which is the truth table of the reverse-ANF; and, from the reverse-ANF, we compute the B-representation and, therefore, the minterms representation of $\{s_{\tau}\}$.

Next, we introduce some results where we study two important properties about Boolean functions such as balancedness and the maximum degree, both obtained from the r-ANF (or equivalently, from the B-representation) and the ANF of the sequence associated with the function.

The following two theorems allow us to know if the r-ANF of a sequence $\{s_\tau\}$ (and the ANF of the function $f$ associated with it) has maximum degree.
These results are based on  Theorem~4 of \cite{Climent2010}, even though we apply this result to sequences and give an alternative proof.

\begin{theorem}\label{thm:max_degree}
Consider $\{s_\tau\}$ a sequence of period $2^t$, with $t$ a positive integer. Then, the ANF associated to $\{s_\tau\}$ has maximum degree if and only if $w(\vec{c}_{\mu_f})$ is odd.    
\end{theorem}
\begin{proof}
 Suppose that we have a sequence $\{s_\tau\}$ of period $2^t$ whose ANF 
 has maximum degree, that is, 
it has the term $x_1x_2\cdots x_{t}$ and, therefore, the vector $\vec{m}_{\mu_f}$ has 1 in the last position.
Remember that all the binomial sequences have $1$ in the last position. Due to the fact that the rows of the matrix $H_t$ are composed of the first $2^t-1$ binomial sequences, and from expression~\eqref{eq:matricial}, we obtain $\vec{m}_{\mu_f}$ as a linear combination of the rows of $H_t$. Thus, $\{s_{\tau}\}$ is formed by the sum of an odd number of binomial sequences, that is, 
$w(\vec{c}_{\mu_f})$ is odd.

Now, suppose that $w(\vec{c}_{\mu_f})$ is odd. 
In a similar way as in the previous implication,
from expression~\eqref{eq:matricial}, we obtain $\vec{m}_{\mu_f}$ as a linear combination of and odd number of rows of $H_t$.
Therefore,  $\vec{m}_{\mu_f}$ has a 1 in the last position and the ANF has maximum degree.  
\end{proof}

We obtain a similar result for the r-ANF of a sequence.
\begin{theorem}
 Consider $\{s_\tau\}$ a sequence of period $2^t$, with $t$ a positive integer. Then, the r-ANF of $\{s_\tau\}$ has maximum degree if, and only if, $\binom{n}{2^t-1}$ is a member of the r-ANF of $\{s_\tau\}$.    
\end{theorem}  
\begin{proof}
 Consequence of Theorem \ref{thm:max_degree} and the isomorphism properties of \eqref{eq:map}.
\end{proof}

Next theorem provides a necessary condition for a sequence to be balanced.
\begin{theorem} \label{Th:balanced}
Let $\{s_\tau\}$ be a balanced sequence of period $2^t$. Then, the term $\binom{n}{2^t-1}$
is not part of the B-representation. 
  Moreover, the term $\overline{x^{\vec{2^t-1}}}$ is not part of the r-ANF.
\end{theorem}
\begin{proof}
If $\{s_\tau\}$, with period $2^t$, is balanced, then its weight is $2^{t-1}$.
From equation~(\ref{eq:binomial_matricial}), we have that the vector $\vec{c}_{\mu_f}$ is the result of the sum of $2^{t-1}$ binomial sequences, determined by the components of the vector $(s_0,s_1,\ldots,s_{2^t-1})$. We know that all the binomial sequences has $1$ in the last position. Therefore, the last component of $\vec{c}_{\mu_f}$ is zero. It means that the B-representation of $\{s_\tau\}$ does not have the term $ \binom{n}{2^t-1} $. That is, as $c_{2^t-1}=0$, then $\overline{\vec{x}^{2^t-1}}$ is not part of the r-ANF of $\{s_\tau\}$.
\end{proof}

As a consequence of the previous results, we can deduce the following corollary.
\begin{corollary}
The linear complexity $LC$ of a balanced sequence with period $2^{t}$  satisfies $LC < 2^t$.
\end{corollary}



\subsection{Relation between the representations of a sequence}
 
In Theorem~\ref{th:bijection}, we show the relation between between the r-ANF and the B-representation of a sequence.
We can obtain directly one representation from the other one.
Moreover, in Section~\ref{BR_ANF}, we analyse the relation between the ANF and the B-representation of a sequence, obtaining that the vector $\vec{m}_{\mu_f}$ can be considered as the sequence. This connection allows us to obtain the r-ANF of a sequence $\{s_\tau\}$ directly from the sequence. 

From  equation~\eqref{eq:binomial_matricial}
and Theorem~\ref{th:bijection}, we can obtain
\begin{equation*} \label{eq:Boolean_B_rep}
\vec{c}_{\mu_f}= (s_0,s_1,\ldots,s_{2^t-1}) \cdot H_{t}\; \text{mod}\; 2.
\end{equation*}
We can rewrite this equation in terms of the ANF as follows
\begin{equation} \label{eq:Boolean_ANF_rep}
\vec{c}_{\mu_f}= \vec{m}_{\mu_f}\cdot H_{t}\; \text{mod}\; 2;
\end{equation}
that is, we can obtain the B-representation from the ANF, and vice versa. Moreover, we can also obtain the r-ANF from the ANF and vice versa.
We check this relation from the following example.
\begin{example}
Consider again the Example~\ref{ex:BR_ANF}. We have that 
$$\vec{m}_{\mu_f}=(0, 0, 0, 1, 1, 1, 0, 0, 1, 0, 0, 1, 1, 0, 1, 1).$$
Applying equation~(\ref{eq:Boolean_ANF_rep}), we obtain that 
$$\vec{c}_{\mu_f}=(0,0,0,1,1,0,1,1,1,1,1,1,1,0,0,0),$$ 
that is, $b(\{s_\tau\})=\{3,4,6,7,8,9,10,11,12\}$. Therefore, the r-ANF of $\{s_\tau\}$ is
\begin{align*}
\mathcal{B}(\{s_\tau\})& =\overline{x^3}+\overline{x^4}+\overline{x^6}+\overline{x^7}+\overline{x^{8}}+\overline{x^{9}}+\overline{x^{10}}+\overline{x^{11}}+\overline{x^{11}}\\
& = x_2x_1+x_3+x_3x_2+x_3x_2x_1+x_4+x_4x_1+x_4x_2+x_4x_2x_1+x_4x_3.
\end{align*}
\end{example}

Next result establishes that a balanced Boolean function cannot have maximum degree.
Remember that a balanced function is determined by its truth table $\vec{\xi}_f$, and, equivalently, through the vector $\vec{c}_{\mu_f}$.
\begin{theorem}\label{th:maxdegree}
 The ANF of a balanced Boolean function of $t$ variables cannot include the maximum monomial $x_1x_2\cdots x_t$, that is, it does not have maximum degree.
 \end{theorem}

 \begin{proof}
    Consequence of Theorem~\ref{thm:max_degree}.
 \end{proof}

\begin{example}
Consider a balanced Boolean function of $4$ variables whose truth table is
\[
\vec{\xi}_f=(1,0,0,1,1,1,0,0,1,0,1,0,1,1,0,0).
\]
Applying equation~(\ref{eq:TT2ANF}), we obtain that
\[
\vec{m}_{\mu_f}=(1,1,1,0,0,1,0,0,0,0,1,0,0,0,1,0),
\]
so, its ANF is
\begin{align*}
f(\vec{x})& =x^{\vec{0}}+x^{\vec{1}}+x^{\vec{2}}+x^{\vec{5}}+x^{\vec{10}}+x^{\vec{14}} = 1+x_4+x_3+x_2x_4+x_1x_3+x_1x_2x_3,
\end{align*}
which has not the maximum degree. 

If we compute the r-ANF, as the vector $\vec{\xi}_f$ is equivalent to the vector $\vec{c}_{\mu_f}$, we have that $b(\{s_\tau\})=\{0,3,4,5,8,10,12,13\}$. The element $2^4-1=15$ is not part of the B-representation, then from Theorem~\eqref{thm:max_degree}, the r-ANF
\begin{align*}
\mathcal{B}(\{s_\tau\})& =\overline{x^0}+\overline{x^3}+\overline{x^4}+\overline{x^5}+\overline{x^{8}}+\overline{x^{10}}+\overline{x^{12}}+\overline{x^{13}}\\
& = 1+x_1x_2+x_3+x_1x_3+x_4+x_2x_4+x_3x_4+x_1x_2x_4,
\end{align*}
has not maximum degree.
\end{example}
\begin{corollary}
Let $f$ be the Boolean function associated with the sequence $\{s_\tau\}$ of period a power of two, that is, its ANF or r-ANF. If $f$ is balanced, then $f$ has not maximum degree.
\end{corollary}

Next, we present some important results which allows us to know more about the r-ANF through the B-representation.
 Both results follow immediately from the Definition~\ref{Def:Brep}.

 \begin{theorem} \label{Bin_Bool}
The r-ANF of a sequence $\{s_\tau\}$ is 
$\mathcal{B}(\{s_\tau\})=x_j$, 
if and only if $b(\{s_\tau\})=\binom{n}{2^{j-1}}$. 
Particularly, $\mathcal{B}(\{s_\tau\})=1$, 
if and only if the B-representation is $b(\{s_\tau\})=\binom{n}{0}$.
\end{theorem}

 Check, for example, the r-ANF of the sequences $\binom{n}{1}$, $\binom{n}{2}$, $\binom{n}{4}$ and $\binom{n}{8}$ in Table~\ref{tab:bin}.

\begin{theorem}\label{th:2L-1}
The  r-ANF of a sequence $\{s_\tau\}$ is 
$\mathcal{B}(\{s_\tau\})=x_jx_{j-1}\cdots x_2x_1$ if and only if the   B-representation is $b(\{s_\tau\})=\binom{n}{2^{j}-1}$.
\end{theorem}

Check, for example, the r-ANF of the sequences $\binom{n}{3}$, $\binom{n}{7}$, and $\binom{n}{15}$ in Table~\ref{tab:bin}.

Next result allows us to obtain the r-ANF of a sequence $\{s_\tau\}$ from its B-representation.

 \begin{theorem}\label{th:bolinear}
 Given a sequence $\{s_\tau\}$ with  B-representation
 $b(\{s_\tau\})=\sum_{i=0}^{2^t-1}c_i  \binom{n}{i} $,
 we have that  its r-ANF is given by 
 $$\mathcal{B}(\{s_\tau\})=\sum_{i=0}^{2^t-1}c_i\mathcal{B}\left(   \binom{n}{i} \right)$$
 \end{theorem}
 \begin{proof}
 Note that the map considered in equation (\ref{eq:map}) is a linear operator, that is, for any two sequences $\{s_\tau\}$ and $\{u_\tau\}$, both with period power of two, we have that $\phi(\{s_\tau\}+\{u_\tau\})=\phi(\{s_\tau\})+\phi(\{u_\tau\})$.
 Therefore,
 $$
 \mathcal{B} \left( 
 \sum_{i=0}^{2^t-1}c_i \binom{n}{i} \right)=
 \phi \left( 
 \sum_{i=0}^{2^t-1}c_i \binom{n}{i} \right)
 =
 \sum_{i=0}^{2^t-1}c_i\phi\left( \binom{n}{i} \right)
 =
  \sum_{i=0}^{2^t-1}c_i\mathcal{B}\left(  \binom{n}{i} \right). $$
  \end{proof}

 Theorem~\ref{th:bolinear} implies that the application $\phi$ given  in 
(\ref{eq:map}) is an $\mathbb{F}_2$-linear space isomorphism.
Therefore, in order to obtain the r-ANF of a sequence, we can just sum over $\Fset_2$ the r-ANFs of the binomial  sequences involved in the B-representation. 
 As an example, observe that the r-ANF of
 the sequence $ \binom{n}{7}+\binom{n}{6} $ is given by
 $$\mathcal{B}\left( \binom{n}{7}+\binom{n}{6} \right)=
 \mathcal{B}\left(
 \binom{n}{7} 
 \right)+\mathcal{B}\left(  \binom{n}{6} \right)=
 x_3x_2x_1+x_3x_2.$$


\subsection{reverse-ANF of shifted sequences}

It is well known that, when a sequence is shifted a finite number of positions, the $B$-representation changes \cite{Cardell2019a}. 
As a consequence, the same happens with the reverse-ANF. 
In this section, we study the r-ANF of shifted versions of a sequence.

 \begin{theorem} \label{Th:shift}
Consider the binomial sequence  $ \binom{n}{2^{j-1}}$ represented by $x_j$, for $j\in\{2,3,\ldots\}$.
If we shift cyclically such a
sequence one bit to the left, then the r-ANF of the shifted version is
$x_1x_2\cdots x_{j-1}+x_j$.
When $j=1$, we have that the r-ANF of  its shifted version is $1+x_1$.
\end{theorem}

\begin{proof}
    Consequence of the isomorphism properties of (\ref{eq:map}), Theorem~\ref{th:2L-1} and Lemma 9 in \cite{Cardell2019a}.
\end{proof}

 \begin{example}
Consider the sequence 
$$\binom{n}{8}=\{0000000011111111\ldots  \}$$
whose r-ANF is $x_4$.
If we shift the sequence one bit to the left, we obtain the following shifted version of the same sequence (see~\cite[Lemma 9]{Cardell2019a} for more details):
$$\binom{n}{8}+\binom{n}{7}=\{0000000111111110 \ldots  \},$$
whose r-ANF is 
$x_1x_2x_3+x_4$.
\end{example}

We can generalize the result given in Theorem~\ref{Th:shift} for sequences whose r-ANFs are monomials.

\begin{theorem}\label{Th:shift2}
Consider the binary sequence  represented by $x_{i_k}x_{i_{k-1}}\cdots x_{i_2} x_{i_1}$ with
$i_k> i_{k-1} > \cdots >i_2> i_1$.
If we shift cyclically such a
sequence one bit to the left, then the r-ANF of the  resultant shifted version is
$x_{i_k}x_{i_{k-1}}\cdots x_{i_3}x_{i_2} + x_{i_k}x_{i_{k-1}}\cdots x_{i_2}x_{i_1}$.
\end{theorem}
 
\begin{proof}
Consequence of the isomorphism properties of (\ref{eq:map}) and Lemma 9 in \cite{Cardell2019a}.
\end{proof}

 \begin{example} \label{ex:shift}
Consider the sequence 
$$\binom{n}{7}=\{0000000100000001\ldots  \}$$
whose r-ANF is $x_3x_2x_1$.
If we shift the sequence one bit to the left, we obtain the following shifted version of the same sequence (see~\cite[Lemma 9]{Cardell2019a} for more details):
$$\binom{n}{7}+\binom{n}{6}=\{0000001000000010\ldots  \},$$
whose r-ANF is 
$x_3x_2x_1+x_3x_2$.
\end{example}


  As a consequence of Theorems~\ref{Th:shift} and \ref{Th:shift2}, we can introduce the following theorem. 

\begin{theorem}\label{th:Borep}
Consider the r-ANF of a sequence 
$\mathcal{B}(\{s_\tau\})=\sum_{i=0}^{2^t-1}c_i \overline{x^{\vec{i}}}.$
If we shift the sequence one bit to the left, we obtain a shifted version of the same sequence with r-ANF given by
$$
\mathcal{B}(\{s_\tau\})=\sum_{i=1}^{2^t-1}c_i \left(\ \overline{x^{\vec{i}}}+\Delta \left(\ \overline{x^{\vec{i}}}\ \right)\right)+c_0.
$$
where
   $$
\Delta\left(\ \overline{x^{\vec{i}}}\ \right)=   \begin{cases}
  x_{i_j}\cdots x_{i_3}x_{i_{2}}     & \text{if } \quad \overline{x^{\vec{i}}} = x_{i_j}\cdots x_{i_{2}}x_{i_{1}}\quad  \text{ for some } i_1<i_2<\cdots < i_j. \\
   x_{j-1}\cdots x_{2}x_1     & \text{ if }  \quad \overline{x^{\vec{i}}} = x_j \quad \text{ for some } j.
  \end{cases}
  $$
  \end{theorem}

\begin{proof}
    Consequence of the isomorphism properties of (\ref{eq:map}) and Theorem 15 in \cite{Cardell2019a}.
\end{proof}

  \begin{example}
  Consider the sequence $\{s_\tau\}=\{01000100\ldots\}$
  represented by  $b(\{s_\tau\})=\binom{n}{3}+\binom{n}{1}$. 
  The r-ANF of this sequence is given by $\mathcal{B}(\{s_\tau\})= x_2x_1+x_1$.
 This sequence has period equal to 4, therefore it has 4 different shifted versions.
The B-representations of the shifted versions are 
  \begin{align*}
b(\{01000100 \ldots\}) & = 
  \binom{n}{3}+\binom{n}{1}, \\
  b(\{10001000 \ldots\}) & =\binom{n}{3}+\binom{n}{2}+\binom{n}{1}+\binom{n}{0},\\
   b(\{00010001 \ldots\}) & =\binom{n}{3}, \\
   b(\{00100010 \ldots\}) & =\binom{n}{3}+\binom{n}{2};\\
  \end{align*}
  and their r-ANFs are
   \begin{align*}
\mathcal{B}(\{01000100 \ldots\}) & = 
  x_2x_1+x_1, \\
  \mathcal{B}(\{10001000 \ldots\}) & =x_2x_1+x_2+x_1+1,\\
   \mathcal{B}(\{00010001 \ldots\}) & =x_2x_1, \\
   \mathcal{B}(\{00100010 \ldots\}) & =x_2x_1+x_2.\\
  \end{align*}
Figure~\ref{fig:rep} shows the process followed  in order to obtain the four r-ANFs using Theorem~\ref{th:Borep}.
  \end{example}
  
  \begin{figure}
\caption{The r-ANFs of the 4 shifted versions of the sequence given by $x_2x_1+x_1$\label{fig:rep}}
\begin{center}
\begin{tikzpicture}[scale=1.5]
\draw [fill=gray!20] (-0.5,-0.25)rectangle (1.6,0.25);
\draw [fill=gray!20] (-0.9,-1)rectangle (1.8,-0.5);
\draw  [fill=gray!20](-1.15,-1.75)rectangle (-0.6,-1.25);
\draw [fill=gray!20] (-1.6,-2.5)rectangle (-0.2,-2);

\draw (-0.2,0) node [scale=1] {$\pmb{x_2x_1}$};
\draw (1.3,0) node [scale=1] {$\pmb{x_1}$};
\draw (1.2,0) node [scale=1] {$\textcolor{red}{}$};

\draw (0.5,0) node {+};

\draw [-latex](-0.2,-0.25)--(-0.6,-0.5);
\draw [-latex](-0.2,-0.25)--(0.2,-0.5);
\draw [-latex](1.3,-0.25)--(1.6,-0.5);
\draw [-latex](1.3,-0.25)--(1,-0.5);

\draw (-0.6,-0.75) node[scale=1] {$\pmb{x_2x_1}$};
\draw (0.2,-0.75) node[scale=1] {$ \pmb{x_2}$};
\draw (1,-0.75) node[scale=1] {$\pmb{x_1}$};
\draw (1.6,-0.75) node[scale=1] {$\pmb{1}$};

\draw [-latex](-0.6,-1)--(-0.8,-1.25);
\draw [-latex](-0.6,-1)--(-0.4,-1.25);

\draw [-latex](0.2,-1)--(0,-1.25);
\draw [-latex](0.2,-1)--(0.4,-1.25);

\draw [-latex](1,-1)--(0.8,-1.25);
\draw [-latex](1,-1)--(1.2,-1.25);

\draw [-latex](1.6,-1)--(1.6,-1.25);

\draw (-0.85,-1.5) node[scale=1] {$\pmb{x_2x_1}$};
\draw (-0.4,-1.5) node[scale=1] {$\cancel{x_2}$};
\draw (0,-1.5) node[scale=1] {$\cancel{x_2}$};
\draw (0.4,-1.5) node[scale=1] {$\cancel{x_1}$};
\draw (0.8,-1.5) node[scale=1] {$\cancel{x_1}$};
\draw (1.2,-1.5) node[scale=1] {$\cancel{1}$};
\draw (1.6,-1.5) node[scale=1] {$\cancel{1}$};

\draw [-latex](-0.8,-1.75)--(-1.4,-2);
\draw [-latex](-0.8,-1.75)--(-0.4,-2);

\draw (-1.35,-2.25) node[scale=1] {$\pmb{x_2x_1}$};
\draw (-0.4,-2.25) node[scale=1] {$\pmb{x_2}$};

\draw [-latex](-1.4,-2.5)--(-1.6,-2.75);
\draw [-latex](-1.4,-2.5)--(-1.2,-2.75);

\draw [-latex](-0.4,-2.5)--(-0.6,-2.75);
\draw [-latex](-0.4,-2.5)--(-0.2,-2.75);

\draw (-1.6,-3) node[scale=1] {$\pmb{x_2x_1}$};
\draw (-1.2,-3) node[scale=1] {$\cancel{x_2}$};
\draw (-0.6,-3) node[scale=1] {$\cancel{x_2}$};
\draw (-0.2,-3) node[scale=1] {$\pmb{x_1}$};

%

\end{tikzpicture}
\end{center}
\end{figure}

Next, we introduce a result relating the Sierpinski triangle with the shifted versions of a binomial sequence. 
Recall that  if we arrange the binomial coefficients  $\binom{n}{k}$ into rows for successive values of $n = 0, 1, 2, \ldots $, then the generated structure is the Pascal's triangle (see Figure~\ref{fig:pascal}).
 If we color the odd numbers of the Pascal's triangle and shade the even ones, we obtain the Sierpinski's triangle (see Figure~\ref{fig:sierp}).
 Considering the Pascal's triangle modulo 2 (Figure ~\ref{fig:sierp2}), we can observe that the diagonals  are the binomial sequences starting in a different position (here they start with the first 1).

\begin{figure}
\caption{Binomial coefficients arranged as triangles}
\centering
\subfloat[Pascal's triangle \label{fig:pascal}]{
\centering
\begin{tikzpicture}[scale=1.75]
\draw (0,0) node [scale=1.4] {$\binom{0}{0}$};

\draw (-0.3,-0.5) node[scale=1.4] {$\binom{1}{0}$};
\draw (0.3,-0.5) node[scale=1.4] {$\binom{1}{1}$};

\draw (-0.6,-1) node[scale=1.4] {$\binom{2}{0}$};
\draw (0,-1) node[scale=1.4] {$\binom{2}{1}$};
\draw (0.6,-1) node[scale=1.4] {$\binom{2}{2}$};

\draw (-0.9,-1.5) node[scale=1.4] {$\binom{3}{0}$};
\draw (-0.3,-1.5) node [scale=1.4]{$\binom{3}{1}$};
\draw (+0.3,-1.5) node[scale=1.4] {$\binom{3}{2}$};
\draw (0.9,-1.5) node [scale=1.4]{$\binom{3}{3}$};

\draw (-1.2,-2) node[scale=1.4] {$\binom{4}{0}$};
\draw (-0.6,-2) node[scale=1.4] {$\binom{4}{1}$};
\draw (-0,-2) node[scale=1.4] {$\binom{4}{2}$};
\draw (+0.6,-2) node[scale=1.4] {$\binom{4}{3}$};
\draw (1.2,-2) node[scale=1.4] {$\binom{4}{4}$};

\draw (-1.5,-2.5) node[scale=1.4] {$\binom{5}{0}$};
\draw (-0.9,-2.5) node[scale=1.4] {$\binom{5}{1}$};
\draw (-0.3,-2.5) node[scale=1.4] {$\binom{5}{2}$};
\draw (0.3,-2.5) node[scale=1.4] {$\binom{5}{3}$};
\draw (+0.9,-2.5) node[scale=1.4] {$\binom{5}{4}$};
\draw (1.5,-2.5) node[scale=1.4] {$\binom{5}{5}$};
\end{tikzpicture}
}
\qquad
\subfloat[Sierpinski's triangle\label{fig:sierp}]{
\centering
\begin{tikzpicture}[scale=0.7]
\phantom{\draw (0,0) circle (0.5);}

\draw [ color=black, fill=red!20](0,0)--(-0.5,-0.87)--(0.5,-0.87)--cycle;
\draw [color=black, fill=red!20](0.5,-0.87)--(0,-1.74)--(1,-1.74)--cycle;
\draw [color=black, fill=red!20](-0.5,-0.87)--(0,-1.74)--(-1,-1.74)--cycle;
\draw [color=black, fill=red!20](1,-1.74)--(1.5, -2.61)--(0.5,-2.61)--cycle;
\draw [color=black, fill=red!20](-1,-1.74)--(-1.5, -2.61)--(-0.5,-2.61)--cycle;
\draw [color=black, fill=red!20](1.5, -2.61)--(2,-3.48)--(1,-3.48)--cycle;
\draw [color=black, fill=red!20](0.5, -2.61)--(1,-3.48)--(0,-3.48)--cycle;
\draw [color=black, fill=red!20](-0.5, -2.61)--(-1,-3.48)--(0,-3.48)--cycle;
\draw [color=black, fill=red!20](-1.5, -2.61)--(-2,-3.48)--(-1,-3.48)--cycle;
\draw [ color=black, fill=red!20](-2,-3.48)--(-2.5,-4.35)--(-1.5,-4.35)--cycle;
\draw [color=black, fill=red!20](-1.5,-4.35)--(-2,-5.22)--(-1,-5.22)--cycle;
\draw [color=black, fill=red!20](-2.5,-4.35)--(-2,-5.22)--(-3,-5.22)--cycle;
\draw [color=black, fill=red!20](-1,-5.22)--(-0.5, -6.09)--(-1.5,-6.09)--cycle;
\draw [color=black, fill=red!20](-3,-5.22)--(-3.5, -6.09)--(-2.5,-6.09)--cycle;
\draw [color=black, fill=red!20](-0.5, -6.09)--(0,-6.96)--(-1,-6.96)--cycle;
\draw [color=black, fill=red!20](-1.5, -6.09)--(-1,-6.96)--(-2,-6.96)--cycle;
\draw [color=black, fill=red!20](-2.5, -6.09)--(-3,-6.96)--(-2,-6.96)--cycle;
\draw [color=black, fill=red!20](-3.5, -6.09)--(-4,-6.96)--(-3,-6.96)--cycle;
\draw [ color=black, fill=red!20](2,-3.48)--(2.5,-4.35)--(1.5,-4.35)--cycle;
\draw [color=black, fill=red!20](1.5,-4.35)--(2,-5.22)--(1,-5.22)--cycle;
\draw [color=black, fill=red!20](2.5,-4.35)--(2,-5.22)--(3,-5.22)--cycle;
\draw [color=black, fill=red!20](1,-5.22)--(0.5, -6.09)--(1.5,-6.09)--cycle;
\draw [color=black, fill=red!20](3,-5.22)--(3.5, -6.09)--(2.5,-6.09)--cycle;
\draw [color=black, fill=red!20](0.5, -6.09)--(0,-6.96)--(1,-6.96)--cycle;
\draw [color=black, fill=red!20](1.5, -6.09)--(1,-6.96)--(2,-6.96)--cycle;
\draw [color=black, fill=red!20](2.5, -6.09)--(3,-6.96)--(2,-6.96)--cycle;
\draw [color=black, fill=red!20](3.5, -6.09)--(4,-6.96)--(3,-6.96)--cycle;
\draw (0,-0.46) node [scale=0.65] {1};

\draw (0.5,-1.4) node [scale=0.65] {1};
\draw (-0.5,-1.4) node [scale=0.65] {1};

\draw (1,-2.23) node [scale=0.65] {1};\draw (0,-2.23) node [scale=0.65] {2};
\draw (-1,-2.23) node [scale=0.65] {1};

\draw (1.5,-3.12) node [scale=0.65] {1};\draw (0.5,-3.12) node [scale=0.65] {3};
\draw (-1.5,-3.12) node [scale=0.65] {1};\draw (-0.5,-3.12) node [scale=0.65] {3};

\draw (2,-4.01) node [scale=0.65] {1};\draw (1,-4.01) node [scale=0.65] {4};\draw (-1,-4.01) node [scale=0.65] {4};
\draw (-2,-4.01) node [scale=0.65] {1};\draw (0,-4.01) node [scale=0.65] {6};

\draw (2.5,-4.9) node [scale=0.65] {1};\draw (0.5,-4.9) node [scale=0.65] {10};
\draw (-2.5,-4.9) node [scale=0.65] {1};\draw (-0.5,-4.9) node [scale=0.65] {10};
\draw (1.5,-4.9) node [scale=0.65] {5};
\draw (-1.5,-4.9) node [scale=0.65] {5};

\draw (3,-5.79) node [scale=0.65] {1};\draw (-0,-5.79) node [scale=0.65] {20};
\draw (-3,-5.79) node [scale=0.65] {1};\draw (2,-5.79) node [scale=0.65] {6};\draw (-2,-5.79) node [scale=0.65] {6};
\draw (-1,-5.79) node [scale=0.65] {15};
\draw (1,-5.79) node [scale=0.65] {15};

\draw (3.5,-6.68) node [scale=0.65] {1};
\draw (-3.5,-6.68) node [scale=0.65] {1};
\draw (-2.5,-6.68) node [scale=0.65] {7};
\draw (-1.5,-6.68) node [scale=0.65] {21};
\draw (-0.5,-6.68) node [scale=0.65] {35};
\draw (0.5,-6.68) node [scale=0.65] {35};
\draw (1.5,-6.68) node [scale=0.65] {21};
\draw (2.5,-6.68) node [scale=0.65] {7};
\end{tikzpicture}
}
\qquad
\subfloat[Sierpinski's triangle mod 2\label{fig:sierp2}]{
\centering
\begin{tikzpicture}[scale=0.7]
\phantom{\draw (0,0) circle (0.5);}

\draw [ color=black, fill=red!20](0,0)--(-0.5,-0.87)--(0.5,-0.87)--cycle;
\draw [color=black, fill=red!20](0.5,-0.87)--(0,-1.74)--(1,-1.74)--cycle;
\draw [color=black, fill=red!20](-0.5,-0.87)--(0,-1.74)--(-1,-1.74)--cycle;
\draw [color=black, fill=red!20](1,-1.74)--(1.5, -2.61)--(0.5,-2.61)--cycle;
\draw [color=black, fill=red!20](-1,-1.74)--(-1.5, -2.61)--(-0.5,-2.61)--cycle;
\draw [color=black, fill=red!20](1.5, -2.61)--(2,-3.48)--(1,-3.48)--cycle;
\draw [color=black, fill=red!20](0.5, -2.61)--(1,-3.48)--(0,-3.48)--cycle;
\draw [color=black, fill=red!20](-0.5, -2.61)--(-1,-3.48)--(0,-3.48)--cycle;
\draw [color=black, fill=red!20](-1.5, -2.61)--(-2,-3.48)--(-1,-3.48)--cycle;
\draw [ color=black, fill=red!20](-2,-3.48)--(-2.5,-4.35)--(-1.5,-4.35)--cycle;
\draw [color=black, fill=red!20](-1.5,-4.35)--(-2,-5.22)--(-1,-5.22)--cycle;
\draw [color=black, fill=red!20](-2.5,-4.35)--(-2,-5.22)--(-3,-5.22)--cycle;
\draw [color=black, fill=red!20](-1,-5.22)--(-0.5, -6.09)--(-1.5,-6.09)--cycle;
\draw [color=black, fill=red!20](-3,-5.22)--(-3.5, -6.09)--(-2.5,-6.09)--cycle;
\draw [color=black, fill=red!20](-0.5, -6.09)--(0,-6.96)--(-1,-6.96)--cycle;
\draw [color=black, fill=red!20](-1.5, -6.09)--(-1,-6.96)--(-2,-6.96)--cycle;
\draw [color=black, fill=red!20](-2.5, -6.09)--(-3,-6.96)--(-2,-6.96)--cycle;
\draw [color=black, fill=red!20](-3.5, -6.09)--(-4,-6.96)--(-3,-6.96)--cycle;
\draw [ color=black, fill=red!20](2,-3.48)--(2.5,-4.35)--(1.5,-4.35)--cycle;
\draw [color=black, fill=red!20](1.5,-4.35)--(2,-5.22)--(1,-5.22)--cycle;
\draw [color=black, fill=red!20](2.5,-4.35)--(2,-5.22)--(3,-5.22)--cycle;
\draw [color=black, fill=red!20](1,-5.22)--(0.5, -6.09)--(1.5,-6.09)--cycle;
\draw [color=black, fill=red!20](3,-5.22)--(3.5, -6.09)--(2.5,-6.09)--cycle;
\draw [color=black, fill=red!20](0.5, -6.09)--(0,-6.96)--(1,-6.96)--cycle;
\draw [color=black, fill=red!20](1.5, -6.09)--(1,-6.96)--(2,-6.96)--cycle;
\draw [color=black, fill=red!20](2.5, -6.09)--(3,-6.96)--(2,-6.96)--cycle;
\draw [color=black, fill=red!20](3.5, -6.09)--(4,-6.96)--(3,-6.96)--cycle;
\draw (0,-0.46) node [scale=0.65] {1};

\draw (0.5,-1.4) node [scale=0.65] {1};
\draw (-0.5,-1.4) node [scale=0.65] {1};

\draw (1,-2.23) node [scale=0.65] {1};\draw (0,-2.23) node [scale=0.65] {0};
\draw (-1,-2.23) node [scale=0.65] {1};

\draw (1.5,-3.12) node [scale=0.65] {1};\draw (0.5,-3.12) node [scale=0.65] {1};
\draw (-1.5,-3.12) node [scale=0.65] {1};\draw (-0.5,-3.12) node [scale=0.65] {1};

\draw (2,-4.01) node [scale=0.65] {1};\draw (1,-4.01) node [scale=0.65] {0};\draw (-1,-4.01) node [scale=0.65] {0};
\draw (-2,-4.01) node [scale=0.65] {1};\draw (0,-4.01) node [scale=0.65] {0};

\draw (2.5,-4.9) node [scale=0.65] {1};\draw (0.5,-4.9) node [scale=0.65] {0};
\draw (-2.5,-4.9) node [scale=0.65] {1};\draw (-0.5,-4.9) node [scale=0.65] {0};
\draw (1.5,-4.9) node [scale=0.65] {1};
\draw (-1.5,-4.9) node [scale=0.65] {1};

\draw (3,-5.79) node [scale=0.65] {1};\draw (-0,-5.79) node [scale=0.65] {0};
\draw (-3,-5.79) node [scale=0.65] {1};\draw (2,-5.79) node [scale=0.65] {0};\draw (-2,-5.79) node [scale=0.65] {0};
\draw (-1,-5.79) node [scale=0.65] {1};
\draw (1,-5.79) node [scale=0.65] {1};

\draw (3.5,-6.68) node [scale=0.65] {1};
\draw (-3.5,-6.68) node [scale=0.65] {1};
\draw (-2.5,-6.68) node [scale=0.65] {1};
\draw (-1.5,-6.68) node [scale=0.65] {1};
\draw (-0.5,-6.68) node [scale=0.65] {1};
\draw (0.5,-6.68) node [scale=0.65] {1};
\draw (1.5,-6.68) node [scale=0.65] {1};
\draw (2.5,-6.68) node [scale=0.65] {1};
\end{tikzpicture}
}
\end{figure}

First, we need to introduce a minor result regarding the rows of the (binary) Sierpinski triangle. 

\begin{lemma}\label{lem:Sierp}
Consider $d^n=(d^n_0,d^n_1, \ldots, d^n_{n})$, for $n=0, 1, \ldots$,  the $n$-th row of the binary Sierpinski triangle (see Figure~\ref{fig:sierp2}). Then,
    \begin{enumerate}

        \item The first and the last elements of $d^n$ are equal to 1, i.e., $d^n_0=d^n_{n}=1$.
        \item It is possible to generate the elements in the $(n+1)$-th row,  denoted by $d^{n+1}$, using the elements in $d^n$, in the following way:   $d^{n+1}_{k+1}=d^n_k+d^n_{k+1}$, for $k=1, \ldots, n-1$.
          
    \end{enumerate}
\end{lemma}
\begin{proof}
    \begin{enumerate}
        \item The first and last elements are $\binom{n}{0}=\binom{n}{n}=1$.
        \item Consequence of the following combinatorial property:
        $$
        \binom{n}{k}+\binom{n}{k+1}=\binom{n+1}{k+1}
        $$
    \end{enumerate}
\end{proof}

\begin{theorem}\label{thm:Sierp}
Consider the binomial sequence $\binom{n}{k}$  with period $T=2^t$ and consider $r$ a positive integer such  that $r < T$.
If we shift such sequence  $r$ positions to the left, then the r-ANF of the resultant shifted version  is given by 
\begin{equation}\label{eq:Sierp:sum}
     \sum_{i=0}^{\min\{r,k\}} d_{r-i}^r \phi \left( \binom{n}{k-i}  \right)
\end{equation}
 
  where $d^r=(d_0^r,d_1^r, \ldots, d_{r}^r)$ is the $r$-th row of the binary Sierpinski triangle (see Figure~\ref{fig:sierp2}). 
\end{theorem}

\begin{proof}





  Let us start with the binomial sequence $\binom{n}{k}$.
If we shift this sequence one position to the left we obtain the sequence $\binom{n}{k-1}+\binom{n}{k}$ (see \cite{Cardell2019a} for more details).
This means that the r-ANF of this shifted version  is
$d_0^1\phi\left(\binom{n}{k-1}\right)+d_1^1\phi\left(\binom{n}{k}\right)$, where $d^1=(d^1_0,d^1_1)=(1,1)$ is the 1st
  row of Sierpinski (Note that the 0-th row is $d^0=(1)$).

Assume now that after $r$ shifts (with $r < k $), the B-representation is 
$$
\sum^{r}_{i=0} d^r_{r-i} \binom{n}{k-i}= d_0^{r}\binom{n}{k-r}+d_1^{r}\binom{n}{k-r+1}+\cdots + d_{r-1}^{r}\binom{n}{k-1}+d_r^{r}\binom{n}{k}
$$ 
and the corresponding r-ANF is 
$$\mathcal{B} \left(\sum^{r}_{i=0} d^r_{r-i}   \binom{n}{k-i}\right)=
\sum^{r}_{i=0} d^r_{r-i} \mathcal{B} \left(  \binom{n}{k-i}\right).$$
Now, if we shift this $r$-shifted version one position to the left, we obtain that the B-representation is given by  
\begin{equation}\label{eq:r:Sierp}
    \sum^{r}_{i=0} d^r_{r-i} \left(\binom{n}{k-i-1}+\binom{n}{k-i}\right)=  
d_0^r\binom{n}{k-r-1} + \sum_{i=0}^{r-1} (d^r_{r-i}+d_{r-i+1}^r) \binom{n}{k-i} + d_r^r\binom{n}{k} 
\end{equation}
(see~\cite{Cardell2019a} for more details).
According to Lemma~\ref{lem:Sierp}, we have that $d_0^r=d_0^{r+1}=1$, $d_r^r=d_{r+1}^{r+1}=1$ and $d_{r-i}^r+d_{r-i+1}^r=d_{r-i+1}^{r+1}$.
Therefore, expression (\ref{eq:r:Sierp}) is equal to
$$
\sum_{i=0}^{r+1} d_{r+1-i}^{r+1}\binom{n}{k-i}
$$
and  the corresponding r-ANF for $r+1$ shifts is given by
$$
\mathcal{B} \left(\sum_{i=0}^{r+1} d_{r+1-i}^{r+1}\binom{n}{k-i}  \right)=\sum_{i=0}^{r+1} d_{r+1-i}^{r+1}\mathcal{B} \left(\binom{n}{k-i}\right),
$$
which proofs that the expression (\ref{eq:Sierp:sum}) is correct for $r<k$.

When $r=k$, the r-ANF has the following form 
$$\mathcal{B} \left( \sum_{i=0}^{k} d^k_{k-i}   \binom{n}{k-i}   \right) =
\sum_{i=0}^{k} d_{k-i}^{k} \mathcal{B} \left( \binom{n}{k-i}   \right).$$  
Note that, in this case, the binomial sequence $\binom{n}{0}$ appears for the first time in the B-representation of the shifted version; this means that the term  1 appears for the first time  in the r-ANF. 
It is worth mentioning that the binomial sequence $\binom{n}{0}$ is invariant for translations, i.e., it remains the same after shifts. 
Therefore, if we shift the binomial sequence   $r=k+1$ position to the left, the B-representation has the following form
\begin{equation}\label{eq:d:Sierp:k}
   \sum_{i=0}^{k-1} d_{k-i}^{k}   \left( \binom{n}{k-i-1}+\binom{n}{k-i}   \right) + \binom{n}{0} =
   \sum_{i=1}^{k}(d^k_{k-i}+d^k_{k-i+1})\binom{n}{k-i}+d_k^ k\binom{n}{k}.
\end{equation}
Recalling that $d^{k+1}_{k+1}=d_k ^k=1$ and $d^k_{k-i}+d^k_{k-1-i}=d^{k+1}_{k+1-i}$, we have that the expression in~(\ref{eq:d:Sierp:k}) is equal to
$$
 \sum_{i=0}^{k} 
 d_{k+1-i}^{k+1} \binom{n}{k-i}
.$$  
Therefore, the r-ANF in this case ($r=k+1$) is
$$
\mathcal{B} \left( \sum_{i=0}^{k} 
 d_{k+1-i}^{k+1} \binom{n}{k-i}\right)
 =
  \sum_{i=0}^{k} 
 d_{r-i}^{r} \mathcal{B} \left(\binom{n}{k-i}\right)
.$$  
Using the same argument for any $k+1<r<T$, the theorem follows. 


\end{proof}
\begin{example}
    Consider the binomial sequence 
    $$ \binom{n}{k}= \binom{n}{5} =
    \{
       0   0   0   0   0   1   0   1 \ldots
    \}
    $$   
    whose r-ANF is $x_3x_1$. 
    Assume we shift this sequence $r= 4$   positions to the left, then the shifted version is given by $\{
   0 1 0 1 0 0 0 0\ldots
    \}$.
     The 4th row of the Sierpinski triangle is $d^4=(1,0,0,0,1)$.
     Therefore the r-ANF
     of the shifted version is given by     
     $$
\sum_{i=0}^{4}d_{4-i}^4\mathcal{B} \left( \binom{n}{5-i}  \right)=
\mathcal{B}\left(  \binom{n}{5} \right) + \mathcal{B}\left(   \binom{n}{1}\right)=x_3x_1+x_1.
$$

Now, assume we shift the sequence $r=6$ positions to the left, then the shifted version is given by $\{
      0 1 0 0 0 0 0 1\ldots
    \}$.
The 6th row of the triangle  is $d^6=(1,0,1,0,1,0,1)$. 
     Therefore the r-ANF
     of the shifted version is given by    
     $$
\sum_{i=0}^{5}d_{6-i}^6\ 
\mathcal{B} \left( \binom{n}{5-i}  \right)=
\mathcal{B} \left(  \binom{n}{5} \right) + \mathcal{B} \left(  \binom{n}{3} \right)+\mathcal{B} \left(   \binom{n}{1}\right)=x_3x_1+x_2x_1+x_1.
$$
In this case, since $r>k$, we only used the   last $k+1=6$ elements of row $d^6=(1,\textcolor{red} {0,1,0,1,0,1})$.
Check Figure~\ref{fig:n5} to understand more deeply the relation between the shifted versions and the Sierpisnki triangle. 
\end{example}

\begin{figure}
\caption{Binomial representations of $ \bi{n}{5} $ and the Sierpinski's triangle\label{fig:n5}}
\centering
\subfloat[Binomial representations of $ \bi{n}{5} $ \label{fig:4a}]{
\begin{tikzpicture}[scale=1.5]
\phantom{\draw (-2,0)--(0,0);}
\draw [fill=gray!20] (-0.4,-0.25)rectangle (0,0.25);
\draw [fill=gray!20] (-0.8,-1)rectangle (0.4,-0.5);
\draw  [fill=gray!20](-1,-1.75)rectangle (-0.6,-1.25);
\draw  [fill=gray!20](0.2,-1.75)rectangle (0.6,-1.25);
\draw [fill=gray!20] (-1.25,-2.5)rectangle (0.85,-2);
\draw [fill=gray!20] (-1.7,-3.25)rectangle (-1.3,-2.75);

\draw [fill=gray!20] (-0.9,-4.75)rectangle (0.1,-4.25);
\draw [fill=gray!20] (1.3,-4.75)rectangle (1.9,-4.25);

\draw [fill=gray!20] (1.1,-3.25)rectangle (1.5,-2.75);

\draw [fill=gray!20] (-1.85,-4)rectangle (1.65,-3.5);

\draw [fill=gray!20] (-2.3,-5.5)rectangle (2.1,-5);


\draw (-0.2,0) node [scale=1] {$\textcolor{red}{\binom{n}{5}}$};
\draw (1.2,0) node [scale=1] {$\textcolor{red}{}$};

\draw [-latex](-0.2,-0.25)--(-0.6,-0.5);
\draw [-latex](-0.2,-0.25)--(0.2,-0.5);

\draw (-0.6,-0.75) node[scale=1] {$\textcolor{red}{\binom{n}{4}}$};
\draw (-0.2, -0.75) node [scale=0.8] {$\textcolor{red}{+}$};
\draw (0.2,-0.75) node[scale=1] {$\textcolor{red}{\binom{n}{5}}$};

\draw [-latex](-0.6,-1)--(-0.8,-1.25);
\draw [-latex](-0.6,-1)--(-0.4,-1.25);

\draw [-latex](0.2,-1)--(0,-1.25);
\draw [-latex](0.2,-1)--(0.4,-1.25);

\draw (-0.8,-1.5) node[scale=1] {$\textcolor{red}{\binom{n}{3}}$};
\draw (-0.2, -1.5) node [scale=0.8]{$\textcolor{red}{+}$};
\draw (-0.4,-1.5) node[scale=0.8] {$\cancel{\binom{n}{4}}$};
\draw (-0.2, -1.5) node [scale=0.8]{$\textcolor{red}{+}$};
\draw (0,-1.5) node[scale=0.8] {$\cancel{\binom{n}{4}}$};
\draw (-0.2, -1.5) node [scale=0.8]{$\textcolor{red}{+}$};
\draw (0.4,-1.5) node[scale=1] {$\textcolor{red}{\binom{n}{5}}$};

\draw [-latex](-0.8,-1.75)--(-1.05,-2);
\draw [-latex](-0.8,-1.75)--(-0.55,-2);

\draw [-latex](0.4,-1.75)--(0.15,-2);
\draw [-latex](0.4,-1.75)--(0.65,-2);

\draw (-1.05,-2.25) node[scale=1] {$\textcolor{red}{\binom{n}{2}}$};
\draw (-0.8, -2.25) node [scale=0.8] {$\textcolor{red}{+}$};
\draw (-0.55,-2.25) node[scale=1] {$\textcolor{red}{\binom{n}{3}}$};
\draw (-0.2, -2.25) node [scale=0.8] {$\textcolor{red}{+}$};
\draw (0.15,-2.25) node[scale=1] {$\textcolor{red}{\binom{n}{4}}$};
\draw (0.4, -2.25) node [scale=0.8] {$\textcolor{red}{+}$};
\draw (0.65,-2.25) node[scale=1] {$\textcolor{red}{\binom{n}{5}}$};

\draw [-latex](-1.05,-2.5)--(-1.45,-2.75);
\draw [-latex](-1.05,-2.5)--(-1.05,-2.75);

\draw [-latex](-0.55,-2.5)--(-0.35,-2.75);
\draw [-latex](-0.55,-2.5)--(-0.65,-2.75);

\draw [-latex](0.15,-2.5)--(0.15,-2.75);
\draw [-latex](0.15,-2.5)--(0.45,-2.75);

\draw [-latex](0.65,-2.5)--(1.2,-2.75);
\draw [-latex](0.65,-2.5)--(0.90,-2.85);

\draw (-1.5,-3) node [scale=1] {$\textcolor{red}{\binom{n}{1}}$};
\draw (-1.1,-3) node[scale=0.8] {$\cancel{\binom{n}{2}}$};
\draw (-0.9,-3) node [scale=0.8] {$\textcolor{red}{+}$};
\draw (-0.7,-3) node[scale=0.8] {$\cancel{\binom{n}{2}}$};
\draw (-0.5, -3) node [scale=0.8] {$\textcolor{red}{+}$};
\draw (-0.3,-3) node[scale=0.8] {$\cancel{\binom{n}{3}}$};
\draw (-0.1, -3) node [scale=0.8] {$\textcolor{red}{+}$};
\draw (0.1,-3) node[scale=0.8] {$\cancel{\binom{n}{3}}$};
\draw (0.3, -3) node [scale=0.8] {$\textcolor{red}{+}$};
\draw (0.5,-3)node[scale=0.8] {$\cancel{\binom{n}{4}}$};
\draw (0.7, -3) node [scale=0.8] {$\textcolor{red}{+}$};
\draw (0.9,-3) node[scale=0.8] {$\cancel{\binom{n}{4}}$};
\draw (1.3,-3) node[scale=1] {$\textcolor{red}{\binom{n}{5}}$};

\draw [-latex](-1.5,-3.25)--(-1.7,-3.5);
\draw [-latex](-1.5,-3.25)--(-1.3,-3.5);

\draw [-latex](1.3,-3.25)--(1.1,-3.5);
\draw [-latex](1.3,-3.25)--(1.5,-3.5);

\draw (-1.65,-3.75) node [scale=1] {$\textcolor{red}{\binom{n}{0}}$};
\draw (-1.35,-3.75) node [scale=0.8] {$\textcolor{red}{+}$};
\draw (-1.05,-3.75) node[scale=1] {$\textcolor{red}{\binom{n}{1}}$};
\draw (-0.1, -3.75) node [scale=1] {$\textcolor{red}{+}$};
\draw (0.85,-3.75) node[scale=1] {$\textcolor{red}{\binom{n}{4}}$};
\draw (1.15, -3.75) node [scale=0.8] {$\textcolor{red}{+}$};
\draw (1.45,-3.75) node[scale=1] {$\textcolor{red}{\binom{n}{5}}$};

\draw [-latex](-1.65,-4)--(-1.85,-4.25);
\draw [-latex](-1.65,-4)--(-1.45,-4.25);

\draw [-latex](-1.05,-4)--(-1.25,-4.25);
\draw [-latex](-1.05,-4)--(-0.85,-4.25);

\draw [-latex](0.85,-4)--(0.65,-4.3);
\draw [-latex](0.85,-4)--(0.1,-4.25);

\draw [-latex](1.45,-4)--(1.25,-4.25);
\draw [-latex](1.45,-4)--(1.65,-4.25);

\draw (-1.9,-4.5) node[scale=0.8] {$\cancel{\binom{n}{0}}$};
\draw (-1.6,-4.5) node [scale=0.8] {$\textcolor{red}{+}$};
\draw (-1.3,-4.5) node[scale=0.8] {$\cancel{\binom{n}{0}}$};
\draw (-1,-4.5) node [scale=0.8] {$\textcolor{red}{+}$};
\draw (-0.7,-4.5) node[scale=1] {$\textcolor{red}{\binom{n}{1}}$};
\draw (-0.4, -4.5) node [scale=0.8] {$\textcolor{red}{+}$};
\draw (-0.1,-4.5) node[scale=1] {$\textcolor{red}{\binom{n}{3}}$};
\draw (0.2, -4.5) node [scale=0.8] {$\textcolor{red}{+}$};
\draw (0.5,-4.5)node[scale=0.8] {$\cancel{\binom{n}{4}}$};
\draw (0.8, -4.5) node [scale=0.8] {$\textcolor{red}{+}$};
\draw (1.1,-4.5) node[scale=0.8] {$\cancel{\binom{n}{4}}$};
\draw (1.4, -4.5) node [scale=0.8] {$\textcolor{red}{+}$};
\draw (1.7,-4.5) node[scale=1] {$\textcolor{red}{\binom{n}{5}}$};

\draw [-latex](-0.7,-4.75)--(-2,-5);
\draw [-latex](-0.7,-4.75)--(-1.2,-5);

\draw [-latex](-0.1,-4.75)--(-0.4,-5);
\draw [-latex](-0.1,-4.75)--(0.2,-5);

\draw [-latex](1.7,-4.75)--(1.2,-5);
\draw [-latex](1.7,-4.75)--(1.9,-5);

\draw (-2.1,-5.25) node [scale=1] {$\textcolor{red}{\binom{n}{0}}$};
\draw (-1.7,-5.25) node [scale=0.8] {$\textcolor{red}{+}$};
\draw (-1.3,-5.25) node[scale=1] {$\textcolor{red}{\binom{n}{1}}$};
\draw (-0.9,-5.25) node [scale=0.8] {$\textcolor{red}{+}$};
\draw (-0.5,-5.25) node[scale=1] {$\textcolor{red}{\binom{n}{2}}$};
\draw (-0.1,-5.25) node [scale=0.8] {$\textcolor{red}{+}$};
\draw (0.3,-5.25) node[scale=1] {$\textcolor{red}{\binom{n}{3}}$};
\draw (0.7, -5.25) node [scale=0.8] {$\textcolor{red}{+}$};
\draw (1.1,-5.25) node[scale=1] {$\textcolor{red}{\binom{n}{4}}$};
\draw (1.5, -5.25) node [scale=0.8] {$\textcolor{red}{+}$};
\draw (1.9,-5.25) node[scale=1] {$\textcolor{red}{\binom{n}{5}}$};


\end{tikzpicture}
}
\subfloat[First 8 rows of the Sierpinski triangle\label{fig:4b}]{
\begin{tikzpicture}[scale=1.5]
\phantom{\draw (-2,0)--(0,0);}
\phantom{\draw (0.5,0)--(1.5,0);}
\draw [fill=gray!20] (-0.4,-0.25)rectangle (0,0.25);
\draw [fill=gray!20] (-0.8,-1)rectangle (0.4,-0.5);
\draw  [fill=gray!20](-1,-1.75)rectangle (-0.6,-1.25);
\draw  [fill=gray!20](0.2,-1.75)rectangle (0.6,-1.25);
\draw [fill=gray!20] (-1.25,-2.5)rectangle (0.85,-2);
\draw [fill=gray!20] (-1.5,-3.25)rectangle (-1.1,-2.75);
\draw [fill=gray!20] (0.8,-3.25)rectangle (1.2,-2.75);

\draw [fill=gray!20] (-1.8,-4)rectangle (-0.8,-3.5);
\draw [fill=gray!20] (0.5,-4)rectangle (1.5,-3.5);

\draw [fill=gray!20] (-2.1,-4.75)rectangle (-1.7,-4.25);
\draw [fill=gray!20] (-1,-4.75)rectangle (-0.6,-4.25);

\draw [fill=gray!20] (0.2,-4.75)rectangle (0.6,-4.25);
\draw [fill=gray!20] (1.4,-4.75)rectangle (1.8,-4.25);

\draw [fill=gray!20] (-2.4,-5.5)rectangle (2.1,-5);

\draw (-0.2,0) node [scale=1] {$\textcolor{red}{1}$};
\draw (1.2,0) node [scale=1] {$\textcolor{red}{}$};
 
\draw (-0.6,-0.75) node[scale=1] {$\textcolor{red}{1}$};
\draw (0.2,-0.75) node[scale=1] {$\textcolor{red}{1}$};

\draw (-0.8,-1.5) node[scale=1] {$\textcolor{red}{1}$};
\draw (-0.2,-1.5) node[scale=1] {$0$};
\draw (0.4,-1.5) node[scale=1] {$\textcolor{red}{1}$};
 
\draw (-1.05,-2.25) node[scale=1] {$\textcolor{red}{1}$};
\draw (-0.55,-2.25) node[scale=1] {$\textcolor{red}{1}$};

\draw (0.15,-2.25) node[scale=1] {$\textcolor{red}{1}$};
\draw (0.65,-2.25) node[scale=1] {$\textcolor{red}{1}$};

\draw (-1.3,-3) node[scale=1] {$\textcolor{red}{1}$};
\draw (-0.8,-3) node[scale=1] {$0$};
\draw (-0.2,-3) node[scale=1] {$0$};
\draw (0.4,-3) node[scale=1] {$0$};
\draw (1,-3) node[scale=1] {$\textcolor{red}{1}$};

\draw (-1.6,-3.75) node[scale=1] {$\textcolor{red}{1}$};
\draw (-1.1,-3.75) node[scale=1] {$\textcolor{red}{1}$};
\draw (-0.5,-3.75) node[scale=1] {$0$};
\draw (0.1,-3.75) node[scale=1] {$0$};
\draw (0.7,-3.75) node[scale=1] {$\textcolor{red}{1}$};
\draw (1.3,-3.75) node[scale=1] {$\textcolor{red}{1}$};

\draw (-1.9,-4.5) node[scale=1] {$\cancel{\textcolor{red}{1}}$};
\draw (-1.4,-4.5) node[scale=1] {$0$};
\draw (-0.8,-4.5) node[scale=1] {$\textcolor{red}{1}$};
\draw (-0.2,-4.5) node[scale=1] {$0$};
\draw (0.4,-4.5) node[scale=1] {$\textcolor{red}{1}$};
\draw (1,-4.5) node[scale=1] {$0$};
\draw (1.6,-4.5) node[scale=1] {$\textcolor{red}{1}$};

\draw (-2.2,-5.25) node[scale=1] {$\cancel{\textcolor{red}{1}}$};
\draw (-1.7,-5.25) node[scale=1] {$\cancel{\textcolor{red}{1}}$};
\draw (-1.1,-5.25) node[scale=1] {$\textcolor{red}{1}$};
\draw (-0.5,-5.25) node[scale=1] {$\textcolor{red}{1}$};
\draw (0.1,-5.25) node[scale=1] {$\textcolor{red}{1}$};
\draw (0.7,-5.25) node[scale=1] {$\textcolor{red}{1}$};
\draw (1.3,-5.25) node[scale=1] {$\textcolor{red}{1}$};
\draw (1.9,-5.25) node[scale=1] {$\textcolor{red}{1}$};
\end{tikzpicture}
 }
\end{figure}

    \begin{corollary}
Assume $ \sum_{i=0}^{2^t-1}c_i\binom{n}{i}$ is the B-representation of $\{s_\tau\}$, then the r-ANF of the sequence cyclically shifted $r$ bits to the left is 
\begin{equation*} 
 \sum_{i=1}^{2^t-1}c_i \sum_{j=0}^{\min\{r,i\}} d_{r-j}^r \mathcal{B} \left( \binom{n}{k-j}  \right),
\end{equation*}
where $d^r$ is the $r$-th row of the binary Sierpinski triangle (see Figure~\ref{fig:sierp2}). 
\end{corollary}

\section{The reverse sequence}\label{reverse}
Now, we are ready to introduce the definition of reverse sequence and its relation with the shifted versions of a sequence. 

\begin{definition}
Given the sequence $\{s_\tau\}$ with period $T$, the reverse sequence $\{s_\tau\}^*$ is a sequence of period $T$ as well, where its $\tau$-th term satisfies the equality $s^*_\tau=s_{T-1-\tau}$ for $\tau=0, 1, \ldots, T-1$.
\end{definition}
For instance, the reverse sequence of the binomial sequence $\bi{n}{5}=\{0\ 0\ 0\ 0\ 0\ 1\ 0\ 1 \}$ is the sequence given by $ \bi{n}{5}^*=\{1\ 0\ 1\ 0\ 0\ 0\ 0\ 0 \}$, which is also a shifted version of $\bi{n}{5}$.

\begin{theorem}\label{theorem:3}
The reverse sequence $\bi{n}{k}^*$ of the binomial sequence $\bi{n}{k}$  is a shifted version   $ \bi{n}{k} $  with shift $k$.
\end{theorem}
\begin{proof}
This is consequence of the following facts:
\begin{enumerate}
\item The first $k$ elements of the sequence $\bi{n}{k}$ are zeros and the $(k+1)$-th element is 1.
\item If the period of the sequence $\bi{n}{k}$ is $T$, then the subsequence $\bi{n}{k}_{k+1\leq n \leq  T-1}$ is symmetric, in the sense that such a subsequence is equal to its own reverse sequence (for more detail  on the properties of the binomial sequences please check Figure 8 in \cite{Cardell2019a}). 
\end{enumerate} 
\end{proof}

As a consequence of the previous theorem and Theorem~\ref{thm:Sierp}, we have the following result.

\begin{corollary}
The B-representation of $\bi{n}{k}^*$ can be obtained by using the $k$-th row of the Sierpinski triangle, denoted by $d^k$, 
as follows:
\[b\left(
\bi{n}{k} ^*\right) = \sum_{i=0}^{k} d^k_{i} \; \binom{n}{i}.
\]
\end{corollary}

As a consequence of the previous corollary we have that the r-ANF of the sequence
 $ \bi{n}{k}^*$ is given by
 $$
\mathcal{B}\left(\binom{n}{k}^*\right) = \sum_{i=0}^{k} d^k_{i} \mathcal{B}\left( \binom{n}{i} \right).
$$

\begin{example}\label{example:3}
Given the  sequence $ \bi{n}{5} $, if we want to compute both representations of the reverse sequence, then we use the 5th row of the Sierpinski's triangle, that is, 
$d^5=(1,1,0,0,1,1)$.
In fact, we have that
\[
b\left( \bi{n}{5} ^*\right) =\sum_{t=0}^{5} d^5_{i}\; \binom{n}{i}  =   \bi{n}{0}+ \bi{n}{1}+   \bi{n}{4}+ \bi{n}{5}.
\]
$$\mathcal{B} \left(\bi{n}{5} ^*  \right)= \mathcal{B}\left(\bi{n}{0}+ \bi{n}{1}+   \bi{n}{4}+ \bi{n}{5}\right)=
1+x_1+x_3+x_3x_1
$$
It is easy to check that the reverse sequence is a shifted version of the own sequence:
\begin{align*}
\binom{n}{5}: & \ 0\ 0\ 0\ 0\ 0\ 1\ 0\ 1\ \ldots \\
\binom{n}{5}^*= \binom{n}{0}+\binom{n}{1}+\binom{n}{4}+\binom{n}{5} :&\  1\ 0\ 1\ 0\ 0\ 0\ 0\ 0\ \ldots
\end{align*}
\end{example}

 
\begin{theorem} \label{theorem:6}
Consider a sequence with period $2^t$ and B-representation  
$\sum_{k=0}^{2^t-1}c_k\binom{n}{k}$, then 
the B-representation of its reverse sequence is given by
\[
b\left(\left\{\sum_{k=0}^{2^t-1}c_k\binom{n}{k}\right\}^*\right)= \sum_{k=0}^{2^t-1}c_k\left( \sum_{i=0}^{k} d^k_{i}\binom{n}{i}  \right).
\]
and the r-ANF is
\[
\mathcal{B}\left(\left\{\sum_{k=0}^{2^t-1}c_k\binom{n}{k}\right\}^*\right) = \sum_{k=0}^{2^t-1}c_k\left( \sum_{i=0}^{k} d^k_{i}\mathcal{B}\left(\binom{n}{i} \right) \right).
\]
\end{theorem}
\begin{proof}
This is a consequence of the fact that the reverse version of a given sequence is the linear combination of the reverse sequences of the corresponding binomial sequences in the B-representation of such a sequence, that is:
\[
\left\{\sum_{k=0}^{2^t-1}c_k\binom{n}{k}\right\}^*=\sum_{k=0}^{2^t-1}c_k \bi{n}{k} ^*.
\] 
\end{proof}

\begin{example} \label{example:5}
Consider the sequence $\{s_\tau\}$ with B-representation $ b(\{s_\tau\})=\bi{n}{2}+\bi{n}{5}+\bi{n}{7} $.
We can compute the B-representation of its reverse sequence $\{s_\tau\}^*$ summing up the reverse sequences of the corresponding binomial sequences involved in the B-representation:
\[
\begin{array}{rlc}
\left\{\bi{n}{2}\right\}^*:                                
  & \left\{\bi{n}{0}+\phantom{\bi{n}{1}}+\bi{n}{2}\phantom{+\bi{n}{3}+\bi{n}{4}+\bi{n}{5}+\bi{n}{6}+\bi{n}{7}\ }\right\}
    & \multirow{3}{*}{+}  \\
\left\{\bi{n}{5}\right\}^*:                                
  & \left\{\bi{n}{0}+\bi{n}{1}+\phantom{\bi{n}{2}+\bi{n}{3}}+\bi{n}{4}+\bi{n}{5}\phantom{+\bi{n}{6}+\bi{n}{7}\ }\right\}
    &  \\
\left\{\bi{n}{7}\right\}^*:  
  & \left\{\bi{n}{0}+\bi{n}{1}+\bi{n}{2}+\bi{n}{3}+\bi{n}{4}+\bi{n}{5}+\bi{n}{6}+\bi{n}{7}\right\}
    & \\ \hline
\left\{\bi{n}{2}+\bi{n}{5}+\bi{n}{7}\right\}^*:
  & \left\{\bi{n}{0}+\phantom{\bi{n}{1}+\bi{n}{2}}+\bi{n}{3}+\phantom{\bi{n}{4}+\bi{n}{5}}+\bi{n}{6}+\bi{n}{7}\right\}
    & \\
\end{array}
\]
Therefore, the B-representation and the r-ANF of the reverse sequence will be:
\begin{align*}
 b(\{s_\tau\}^*)=&   \bi{n}{0}+ \bi{n}{3}+\bi{n}{6}+\bi{n}{7}\\\
 \mathcal{B}(\{s_\tau\}^*)= &x_3x_2x_1+x_3x_2+x_2x_1+1.
\end{align*}
It is easy to check that both sequences are reverses:
\[
\begin{array}{rc}
\left\{\bi{n}{2}+\bi{n}{5}+\bi{n}{7} \right\}          :& 0\   0\   1\   1\   0\   1\   1 \  1\\
\left\{\bi{n}{0}+ \bi{n}{3}+\bi{n}{6}+\bi{n}{7}\right\}:& 1\   1\   1 \  0\   1\   1 \  0 \  0\\
\end{array}
\]
\end{example}

\section{Reverse-ANF of generalized sequences}\label{sec:gen}

In \cite{Cardell2020c}, the authors introduced different representations of generalized self-shrunken sequences and in  analysed some criptographic  properties of them through the representations given. In this section, we study this new Boolean representation for sequences, the reverse-ANF, in the particular case of the generalized sequences with the aim to complete the work done in \cite{Cardell2020c}. We present some interesting results which could help us in the research of some open problems related with these sequences. 

Next, we introduce the definition of the generalized sequences and some necessary results required for the understanding of this section.

\color{black}
Consider a PN-sequence $\{u_\tau\}_{\tau\geq 0}$ obtained from a maximal-length LFSR with $L$ stages, an $L$-dimensional  vector $\mathcal{G}=[g_0,g_1,g_2,...,g_{L-1}]\in\Fset_2^L$ and let $\{v_\tau\}_{\tau\geq 0}$ be the sequence defined as:
\begin{equation*} 
 v_\tau=g_0u_\tau+ g_1u_{\tau-1}+g_2 u_{\tau-2} + \cdots + g_{L-1}u_{\tau-L+1} \quad \text{for $\tau \geq 0$}.
\end{equation*}
Now, the decimation rule to generate  new sequences $\{s_j\}_{j\geq 0}$ is given by: 
\begin{equation*}
  \begin{cases}
    \text{If } u_{\tau}=1, & \text{then } s_j = v_\tau, \\
    \text{If } u_{\tau}=0, & \text{then } v_\tau \text{ is discarded}.
  \end{cases}
\end{equation*}

The sequence $\{s_j\}_{j\geq 0}$, denoted by $S(\mathcal{G})$, is called the \textit{generalized self-shrunken sequence},  GSS-sequence or simply \textbf{generalized sequence} associated with $\mathcal{G}$ (see  \cite{Hu2004}); and the sequence generator is called the \textbf{generalized self-shrinking generator} (GSSG).

Note that when $\mathcal{G}$ runs over $\mathbb{F}_{2}^{L} \setminus \{\pmb{0}\}$ we obtain all the shifted versions of $\{u_{\tau}\}_{\tau \geq 0}$.
The set of sequences $\mathcal{F} = \left\{S(\mathcal{G}) \ | \  \mathcal{G}\in\Fset_{2}^L\setminus \{\pmb{0}\}\right\}$ is called the 
\textit{family of generalized sequences} based on the PN-sequence $\{u_\tau\}_{\tau \geq 0}$.
It is easy to check that the family $\mathcal{F} \cup \{S(\vec{0})\}$, where  $S(\vec{0})=\{0\ 0\  0\  0\ 0 \ldots\}$, is an additive group with the operation addition modulo 2 \cite{Hu2004}.
In particular, 
the opposite of any sequence $S(\mathcal{G})$ is the sequence itself.
Moreover, the period of every generalized sequence is a divisor of $2^{L-1}$ (the number of ones in the PN-sequence) and every sequence of this family is balanced except for the identically $1$  sequence and the null sequence \cite[Theorem 1]{Hu2004}.
The  generalized sequences of period $1$ and $2$, that is,
$
\{111111 \ldots \}, \{010101 \ldots \} \text{ and } \{101010 \ldots \},
$
are referred as the trivial sequences of the family of generalized sequences. 
The null sequence is not considered as a generalized sequence.

 
\begin{theorem}
The r-ANF of the trivial sequences in the family of generalized sequences are: 
\begin{equation*}
    \mathcal{B}(\{111111 \ldots \})=1, \quad \mathcal{B}(\{010101 \ldots \})=x_1, \quad \text{ and } \quad \mathcal{B}(\{101010 \ldots \})=1+x_1.
\end{equation*}

\end{theorem}

\begin{example}\label{ex:gen}
Consider the primitive polynomial $p(x)=1+x+x^4$.
In Table \ref{tab:gen} we can find the generalized sequences obtained with $p(x)$ and the corresponding B-representation and r-ANF for each sequence. 
Note that all sequences (except for the trivial ones) contain the monomials $x_3$ or $x_3x_1$, which indicates the linear complexity (6 and 5 in this case).
Notice also that the r-ANF of the first half of the sequences is the same as the inferior half except for the term 1. 
\end{example}
\begin{table}
\caption{B-representation and r-ANF of generalized sequences} \label{tab:gen}
\begin{tabular}{|c|c|c|c|}\hline
$\mathcal{G}$ & \text{Generalized sequence} & B-\text{representation}& r-ANF \\\hline
 0001&  0   1   0   1   0   1   0   1&(1)& $x_1$\\
 0010&  1   0   1   1   0   0   0   1&(0,1,3,4,5)&$1+x_{1}+x_2x_{1}+x_{3}+x_{3}x_{1}$\\
0011&   1   1   1   0   0   1   0   0&(0,3,4,5)&$1+x_2x_{1}+x_{3}+x_{3}x_{1}$\\
 0100&  0   1   1   1   0   0   1   0&(1,2,3,5)&$x_{1}+x_2+x_2x_{1}+x_{3}x_{1}$\\
 0101&  0   0   1   0   0   1   1   1&(2,3,5)&$x_2+x_2x_{1}+x_{3}x_{1}$\\
  0110& 1   1   0   0   0   0   1   1&(0,2,4)&$1+x_2+x_{3}$\\
 0111&  1   0   0   1   0   1   1   0&(0,1,2,4)&$1+x_{1}+x_2+x_{3}$\\
 1000&  1   1   1   1   1   1   1   1&(0)&$1$\\
 1001&  1   0   1   0   1   0   1   0&(0,1)&$1+x_{1}$\\
 1010&  0   1   0   0   1   1   1   0&(1,3,4,5)&$x_{1}+x_2x_{1}+x_{3}+x_3x_{1}$\\
1011&   0   0   0   1   1   0   1   1&(3,4,5)&$x_2x_{1}+x_{3}+x_{3}x_{1}$\\
 1100&  1   0   0   0   1   1   0   1&(0,1,2,3,5)&$1+x_{1}+x_2+x_2x_{1}+x_{3}x_{1}$\\
 1101&  1   1   0   1   1   0   0   0&(0,2,3,5)&$1+x_2+x_2x_{1}+x_{3}x_{1}$\\
   1110&0   0   1   1   1   1   0   0&(2,4)&$x_2+x_{3}$\\
   1111&0   1   1   0   1   0   0   1&(1,2,4)&$x_{1}+x_2+x_{3}$\\\hline
\end{tabular}
\end{table}

Next theorem provides   some important properties of the r-ANF of the generalized sequences.
\color{black}

\begin{theorem} \label{th:gen}
    The r-ANF $\mathcal{B}(\{s_\tau\})$ of any generalized sequence $\{s_\tau\}$ obtained from a primitive polynomial of degree $L$ satisfies the following properties:
    \begin{enumerate}
            \item[a)] \label{th:gena} $\mathcal{B}(\{s_\tau\})$ is composed of $|b(\{s_\tau \})|$ monomials, where  $|b(\{s_\tau \})|$ denotes the support of the B-representation of $ \{s_\tau \}$.
                    \item[b)] \label{th:genb} The maximum sub-index $j$ in the variables $x_j$ of $\mathcal{B}(\{s_\tau\})$ is $L-1$.
        \item[c)] \label{th:genc} $\mathcal{B}(\{s_\tau\})$ does not include the maximum term $x_{L-1}\cdots x_2x_1$. 
        \item [d)] \label{th:gend} There exists another generalized sequence in the same family with r-ANF~$\mathcal{B}(\{s_\tau\})+1$. 
    \end{enumerate}
     
\end{theorem}

\begin{proof}
\begin{enumerate}
   \item[a)] This item follows immediately from the definition of r-ANF.
\item [b)] It is well-known that the period of the generalized sequences is a divisor of $2^{L-1}$ \cite{Hu2004}.
The best case scenario is when the period is exactly $2^{L-1}$.
In this case the last term of the B-representation of the sequence is  $\binom{n}{k}$ with $2^{L-2}\leq k \leq 2^{L-1}-1$.
The binary representation of $k$ 
has length $L-1$,
 therefore, the r-ANF have $L-1$ variables: $x_1$, $x_2$, \ldots, $x_{L-1}$.
 \item [c)]  Except for the trivial sequence
 $\{11111111\ldots\}$, the   generalized sequences are always balanced \cite{Hu2004}.
Therefore,  according to Theorem~\ref{th:maxdegree},
the generalized sequences cannot have maximum degree, i.e., they cannot include the maximum term $x_{L-1}\cdots x_2x_1$. 
\item [d)] According to \cite{Cardell2020c,Cardell2019a}, for any generalized sequence $\{s_\tau\}$ with B-representation $b(\{s_\tau\})$, there exists another generalized sequence in the same family with B-representation $b(\{s_\tau\})+\binom{n}{0}$.
The results follows from the definition of r-ANF. 
\end{enumerate}
  
\end{proof}


As a consequence of item~b) of Theorem~\ref{th:gen}, we can deduce that  the B-representation of a generalized sequence cannot contain the term $\binom{n}{2^{L-1}-1}$, and then the  linear complexity satisfies $LC < 2^{L-1}$.

\begin{example}
    Consider, for example, the generalized sequence $\{s_\tau\}$ in Example~\ref{ex:gen} with r-ANF
$\mathcal{B}(\{s_\tau\})=1+x_{1}+x_2x_{1}+x_{3}+x_{3}x_{1}$
and B-representation
 $b(\{s_\tau\})=(0,1,3,4,5).$
Now, Note that the maximum sub-index in the r-ANF is $L-1=3$ and it does not include the maximum term $x_3x_2x_1$.
Also, the r-ANF is composed of 5 monomials, which coincides with the support of the B-representation. 
It is worth noticing that there exists another generalized sequence in the same family, with r-ANF 
$\mathcal{B}(\{s_\tau\})+1=x_{1}+x_2x_{1}+x_{3}+x_{3}x_{1}$ and B-representation
$(1,3,4,5)$.

\end{example}

From Theorem~\ref{th:gen}, we know that the number of monomials of the r-ANF of a binary sequence is the support of its B-representation. Next result gives us an upper bound of this value for generalized sequences, which is related with the linear complexity of them. 
In \cite{Fuster2020}, authors proved that, for these sequences, $LC\leq 2^{L-1}-(L-2)$.
Note that a similar bound was found in \cite{Blackburn1999,Blackburn1996} for the self-shrinking generator. 
\begin{theorem}  
    Let $\{s_\tau\}$ be a generalized sequence. The number of monomials 
    $\mathcal{N}$ of its r-ANF $\mathcal{B}(\{s_\tau\})$ satisfies  
    \[\mathcal{N}=
    |b(\{s_\tau \})| \leq 2^{L-1}-(L-2).
    \]
\end{theorem}
\begin{proof}
 It is an immediate consequence from the definition B-representation  and the bound of the $LC$ for generalized sequences. The maximum value in the B-representation of a generalized sequence is the term $\left\{\binom{n}{i_k}\right\}$, as we saw in Section~\ref{sec:Brepr}. Moreover, $LC=i_k+1 \leq 2^{L-1}-(L-2)$, that is, $i_k \leq 2^{L-1}-(L-1)$. But, in the worst case, we could have in the B-representation of the sequence all the elements $\left\{\binom{n}{i}\right\}$, for $i=0,\ldots,2^{L-1}-(L-1)$. Therefore, 
$ |b(\{s_\tau \})| \leq 2^{L-1}-(L-2).
    $
\end{proof}

\begin{theorem} \label{th:sop}
    Let $\{s_\tau\}$ be a generalized sequence  of period $2^ {L-1}$  generated by a primitive polynomial   of degree $L$. Then,  $2^{L-2}$
    shifted versions of  $\{s_\tau\}$ have the term $\binom{n}{0}$ in their B-representation, or, in other words, they have the monomial $1$ in their r-ANF.
\end{theorem}
\begin{proof}

 If $\{s_\tau\}$ is balanced, then it will be composed of $2^{L-2}$ zeros and $2^{L-2}$ ones.
This means that half of the shifted versions ($2^{L-2}$ versions) start with 1. 
As a consequence of Theorem~\ref{Bin_Bool}, the binomial representation of those shifted versions includes the term $\binom{n}{0}$, that is, the r-ANF includes the monomial $1$.  
\end{proof}

\begin{table}
    \caption{r-ANF and B-representation of the shifted versions of a generalized sequence of period $8$.}\label{tab:shifted}
\begin{tabular}{|c|c|c|}\hline
\text{Shifted versions}& \text{B-representation} &\text{r-ANF}\\\hline
    1   0   1   1   0   0   0 1 & (0,1,3,4,  5) & $1+x_{1}+x_2x_{1}+x_{3}+x_{3}x_{1}$\\
    0   1   1   0   0   0  1  1 &(1,2,5)&$x_{1}+x_2+x_{3}x_{1}$\\
    1   1   0   0   0  1   1 0  &(0,2,4,5)&$1+x_2+x_{3}+x_{3}x_{1}$\\
    1   0   0   0  1   1 0   1 &(0,1,2,3,5)&$1+x_{1}+x_2+x_2x_{1}+x_{3}x_{1}$\\
    0   0   0   1 1   0   1   1 &(3,4,5)&$x_2x_{1}+x_{3}+x_{3}x_{1}$\\
    0   0   1  1 0   1   1   0 &(2,5)&$x_2+x_{3}x_{1}$\\
    0     1  1 0   1   1 0 0   &(1,2,4,5)&$x_{1}+x_2+x_{3}+x_{3}x_{1}$\\
   1  1 0   1  1 0 0 0&(0,2,3,5)&$1+x_2+x_2x_{1}+x_{3}x_{1}$\\\hline
\end{tabular}
\end{table}


 \begin{example}
In Table~\ref{tab:shifted} we can find the B-representation and the r-ANF of all shifted versions of
$\{ s_\tau\}= \{    10110001 \}$,
one of the  
generalized sequences 
 in Example~\ref{ex:gen}.
Note that the r-ANF of all the shifted versions have the same maximum monomial $x_3x_1$, which corresponds to the term $\binom{n}{5}$ in the B-representation. 
 It is worth noticing that some shifted versions correspond to other generalized sequences in the same family (check Table~\ref{tab:gen}). 
 This usually happens when the degree of the polynomial is small. 
Note that   half of the shifted versions have the term $\binom{n}{0}$ in the B-representation,   as proved in Theorem~\ref{th:sop}. 
\vspace{1em}
\end{example}

\vspace{1cm}

\section{Conclusions}\label{sec:Concl}
In this paper, we define a bijection between the set of Boolean functions and the set of binary sequences of period a power of two. This connection allows us to analyse properties of the binary sequences through Boolean functions. Moreover, we define the reverse-ANF of a sequence, that is, a new representation of binary sequences based on Boolean functions. 
We introduce this new Boolean representation, instead of using the ANF, in order to our representation of a sequence, through Boolean functions, is unique.
We show the relation between the different representations presented in terms of binary sequences and Boolean functions. We study the reverse-ANF in the family of generalized sequences, its relation to the Sierpinski triangle, and the representation of the reverse sequences. 
As future work, we would like to go in depth in the analysis of more important cryptographic properties of Boolean functions, as strict avalanche criterion, correlation immunity, non-linearity, among others, in terms of this r-ANF and how it can be interpreted in function of the binary sequences and other representations. Our aim is to use this connection to try to solve difficult problems in the field of binary sequences using Boolean functions, and vice verse. For instance, an open problem for the generalized sequences, obtained from a primitive polynomial of degree $L$, is to prove that their period is always $2^{L-1}$, except for the trivial sequences with period $1$ and $2$; and that the linear complexity is lower bounded by $2^{L-2}$.

\section*{Acknowledgements}

This work is part of the R+D+i grant P2QProMeTe (PID2020-112586RB-I00)  funded by MICIU/AEI/10.13039/501100011033 (Spain).
The work of the first author was supported by Conselho Nacional de Desenvolvimento Científico    e Tecnológico (CNPq), Brazil, project 405842/2023-6 and by Fapesp with process 2024/05051-7.
The work of the third author was supported by the I+D+i project VIGROB-287 of the Universitat d’Alacant.

\bibliography{biblio}

\end{document}